\documentclass[]{aastex631}

\usepackage{amsmath}

\newcommand{\inlineeqnum}{\refstepcounter{equation}~~\mbox{(\theequation)}}
\shorttitle{Search for periodicities in High Energy AGNs}
\shortauthors{Rueda et al.}
\graphicspath{{./}{figures/}}

\begin{document}

\title{Search for periodicities in High Energy AGNs with a time-domain approach}

\author[0000-0001-9833-7637]{Hector Rueda}

\author{Jean-Francois Glicenstein}

\author{Francois Brun}
\affiliation{IRFU\\
CEA Paris-Saclay\\
F-91191, Gif-sur-Yvette, France}



\begin{abstract}
This paper investigates a new methodology to search for periods in light-curves of high-energy gamma-ray sources such as Active Galactic Nuclei (AGNs). 
High-energy light curves have significant stochastic components, making period detection somewhat challenging.
In our model, periodic terms, drifts of the light-curves and random walk with correlation between flux points due to colored noise are taken into account independently. The parameters of the model are obtained directly from a Markov Chain Monte-Carlo minimization. The time periods found are compared to the output of the publicly available Agatha program. The search method is applied to high-energy periodic AGN candidates from the Fermi-LAT catalogue. The significance of periodic models over pure noise models is discussed. Finally, the variability of the period and amplitude of oscillating terms is studied on the most significant candidates.
\end{abstract}



\section{Introduction} \label{sec:intro}

Systems of binary super-massive black holes (SMBH) are likely to arise in galaxy merging events. These binary black holes (BH) could be responsible for the apparent precession of radio jets \citep{1980Natur.287..307B} and would \textbf{show} detectable periodic modulation of their fluxes.
The typical Keplerian period is 1.6 year for $10^8 \mathrm{M}_{\odot}$ BH located at $10^{16} \mathrm{cm}$ from each other. The interest in binary SMBH has been renewed by the detection of gravitational waves from stellar-mass BH merger events by the LIGO and VIRGO collaboration and the prospect of observation of massive (up to $ 10^6 \mathrm{M}_{\odot}$) BH merger events at upcoming instruments \citep{2021NatRP...3..344B}. 
Black hole candidates are best identified through their radio or high-energy emission. 
One of the most straightforward ways to search for binary black hole candidates is by detecting periodicities on their light curves. 

The Large Area Telescope embarked on the Fermi Gamma Ray Space Telescope ({\it Fermi-LAT}) monitors continuously the High Energy Gamma ray sky. Thousands of high-energy AGNs, mostly blazars, were
detected by the {\it Fermi-LAT} in the 100 MeV-300 GeV passband. The {\it Fermi-LAT} data are publicly available and span a period of more than 12 years (2008-2021). Periodic or "quasi-periodic" AGN light curves have been searched by various groups in the Fermi-LAT AGN sample (a few recent publications are \citet{2021ApJ...907..105Y,2020ApJ...896..134P,2020A&A...634A.120A, 2017ApJ...835..260Z,2017MNRAS.471.3036P}). 

Methods for searching periods belong to 2 broad classes. The first class is spectral-domain searches, which involves building variants 
of the Lomb-Scargle  \citep{2018ApJS..236...16V,1976Ap&SS..39..447L,1982ApJ...263..835S}. 
The other class of searches is performed with time-domain methods, namely gaussian processes \citep{2020ApJ...895..122C, 2021ApJ...907..105Y, 2021ApJ...919...58Z}, CARMA models \citep{2014ApJ...788...33K}, weighted wavelet Z transforms \citep{1996AJ....112.1709F}. Both classes of methods are extensively reviewed and evaluated in \citet{2020ApJS..250....1T}, with the goal of obtaining  power spectral densities from AGNs. \citet{2020ApJS..250....1T} use pure stochastic models and do not discuss the removal of long term linear (or periodic) trends from their light curves. 

This paper uses a time-domain approach similar to the gaussian process approach. The flux signal is analyzed as the sum of a periodic mean signal (with a possible linear trend) and a stochastic baseline. The baseline would be described by a gaussian process for irregularly sampled light curves.
In most time-domain searches for periodicities \citep[e.g.][]{2020ApJ...895..122C, 2021ApJ...919...58Z}, the periodic component is modelled by including a periodic kernel on a stochastic light-curve. While this method provides satisfactory fits to the data, 
we decided to have separate deterministic and 
stochastic components for 2 reasons. The first reason is methodological. Methods for stochastic 
time series used in this paper assume that they are stationary in time. Hence all trends and periodicities have to be fitted before analyzing the stochastic component. The second reason is related to our goal of searching for binary black holes. A periodic component of the light curve could be an "internal" damped oscillations (say from the disk emission) or an external forcing as in the case of binary black holes, which would be strictly periodic, or a combination of both.
In our approach, the deterministic and stochastic components are clearly separated from the start and could thus describe both an external forcing and an internal oscillation.

In this paper, we further concentrate on regularly sampled light curves. The flux can then be modelled by the sum of a periodic flux, a linear trend and an auto-regressive part (an AR model). It would be more general to use ARMA models instead of AR models. 
But AR models have less parameters and they can be more easily related to physical properties of the AGN system such as correlation timescales or eigen-frequencies (see e.g equations \ref{eq:o-u} and \ref{eq:arma_1}). A very strong assumption we make on the stochastic component is that of stationarity. This assumption may be broken during an AGN outburst. The connection between our model and actual physical parameters such as the amplitude of the periodic signal is described in Section \ref{subsec:ts}. Section \ref{sec:methods} describes the integration of this model in a Markov Chain Monte-Carlo (MCMC) minimization program. A sample of several periodic AGN candidates from the 4FGL catalogue are analyzed with this program. Section \ref{sec:data} explains the production of light-curves from these Fermi-LAT candidates. Results from the MCMC search are discussed in Section \ref{sec:results}.

\section{Methods} \label{sec:methods}
This section motivates the choice of the physical model employed for the description of the light-curves. Following the standard practice in time-series analysis, we separate the periodic and stochastic components of the light curve. The stochastic component is colored noise with a typical auto-correlation timescale of a few weeks. The basic idea of our method is to eliminate the correlation between flux measurements at consecutive times using an auxiliary variable (equation \ref{eq:auxvar}).

\subsection{General assumptions} 
\label{subsec:ts}
The following considerations have to be taken into account to build a model of AGN light-curve.
\begin{itemize}
    \item The noise on AGN light curves is known to have a log-normal amplitude distribution (see for instance \citet{2020A&A...634A.120A}). In this paper, we work with the logarithm of the flux (designed below as "lflux").  
    \item light-curve baselines are not stable, and exhibit drifts. In this paper, baseline drifts are taken as linear drift of lflux with time.
    \item  the noise on light-curves could be additive, multiplicative or a combination of both. Noise on a time-scale of a few weeks is modelled as additive colored noise in this work.     
    \item detected periods may not be stable in time. The period instability could be an artefact of a multiplicative source of noise. In this paper, data are analyzed in small time-intervals to take into account a possible drift of the lflux period.
\end{itemize}

The model lflux light-curve $\phi(t)$ is  the sum of a periodic term, a mean lflux $\bar{\phi}$, a linear trend and a stochastic component $\epsilon$. The periodic term can be written (including the mean lflux) as 
\begin{equation}
    \phi_P(t) = \bar{\phi}+\sum_{j} (A_j \cos(\omega_j t) + B_j \sin(\omega_j t))
\label{eq:1}
\end{equation}
The $A_j, B_j$ and $\omega_j$ are assumed to take constant values inside each analysis time interval.

The evolution in time of the stochastic component $\epsilon$ is governed by a stochastic equation. The  Ornstein-Uhlenbeck model is a popular choice for this equation. In this model, $\epsilon$ satisfies to
\begin{equation}
\dot{\epsilon} = -\frac{\epsilon}{\tau} +\frac{\sqrt{D}}{\tau} \zeta(t)
\label{eq:o-u}
\end{equation}
where $\tau$ is the correlation time of noise, $D$ is a constant, $\zeta$ is white noise and the dot is the time derivation. The noise term can also be described by models involving higher order derivatives such as CARMA models \citep{BrockwellDavies}.

Using as an example the Ornstein-Uhlenbeck noise model, the time evolution of the flux satisfies
\begin{eqnarray}
    \phi(t)& = \bar{\phi}+\sum_{j} (A_j \cos(\omega_j t) + B_j \sin(\omega_j t)) + Ct + \epsilon(t)  \label{eq:2_1}\\
    \dot{\epsilon} &= -\frac{\epsilon}{\tau} +\frac{\sqrt{D}}{\tau} \zeta(t)
\label{eq:2_2}
\end{eqnarray}

Equation \ref{eq:2_2} can be formally integrated to
\begin{equation}
\epsilon(t)=\exp{((s-t)/\tau)}\epsilon(s) +w(t-s), \  t\ge s
\label{eq:o-u-i}
\end{equation}
where $w(t-s) \sim N(0, D\times(1-\exp(-2(t-s)/\tau))$ is a Gaussian random variable.

\subsubsection{Regularly sampled light-curves}\label{sec:rlc}
 
If the light-curve is regularly sampled in time, then equation \ref{eq:o-u-i} changes to
\begin{equation}
\epsilon(t_n) = \beta \epsilon(t_{n-1}) + w(t_n-t_{n-1})
\label{eq:arma_1}
\end{equation}
where $\beta$ is independent of time. In other words, $\epsilon(t_n)$ is described by an AR(1) model \citep{BrockwellDavies,ChanTong}.

A simple generalization of equation \ref{eq:arma_1} is the AR(k) model
\begin{equation}
\epsilon(t_n) = \sum_{j=1}^{k} \beta_j \epsilon(t_{n-j}) + w(\delta t).
\label{eq:arma_2} 
\end{equation}
with $\delta t = t_1-t_0.$

The time evolution of $\phi$ and $\epsilon$ is now described by: 
\begin{eqnarray}
    \phi(t_n)& = \bar{\phi}+\sum_{j} (A_j \cos(\omega_j t_n) + B_j \sin(\omega_j t_n)) + Ct_n + \epsilon(t_n)  \label{eq:4_1}\\
    \epsilon(t_n)& = \sum_{j=1}^{k} \beta_j \epsilon(t_{n-j}) + w(\delta t)
\label{eq:4_2}
\end{eqnarray}

$\epsilon$ can be eliminated from equation \ref{eq:4_1} by using the linear combination  
\begin{equation}
z = \phi(t_n) -  \sum_{j=1}^{k} \beta_j \phi(t_{n-j}) \label{eq:auxvar}.\end{equation}
Note that a similar trick also works for 
irregularly sampled light-curves, except that the $\beta_j$ coefficients would depend on time.  

The evolution of $z$ is described by an equation similar to equation \ref{eq:4_1}, except for the $\epsilon$ term which is now replaced by a gaussian random variable.  
\begin{equation}
    z = \bar{\phi'}+\sum_{j} ({A'}_j \cos(\omega_j t_n) + {B'}_j \sin(\omega_j t_n)) + C't_n + w(\delta t) 
    \label{eq:5}
\end{equation}

The physical  $\bar{\phi}, A_j,B_j,C$ variables are obtained from the $\bar{\phi'},{A'}_j,{B'}_j,C'$ variables by the transformation given in appendix A.

The model of equation \ref{eq:5} is closely related to ARIMA models with "exogenous covariates" \citep{2018FrP.....6...80F}, while
its extension to irregular spacing would be a variant of space state models \citep{durbin2001series}. It is however not of common use in high energy astrophysics.
As emphasized in the introduction, it has the advantage of clearly separating the periodic and the stochastic part of the signal. Compared to CARMA based approaches such as those from
\citet{2021ApJ...907..105Y,2018ApJ...863..175G}, it has the potential of identifying multiple periods such as harmonics, giving crucial clues on the underlying physical mechanism of the flux oscillation. Compared to the spectral methods such as Agatha (see Section \ref{subsec:agatha}),  the amplitude of the periodic terms is determined, allowing the study of their evolution with time.
The implementation of 
our model 
in the MCMC minimization is described in further details in section \ref{subsec:mcmc}.  The formalism for the regular sampling is easier to implement in the MCMC search. We therefore focus on the study of regularly sampled light-curves in this paper.

\subsection{MCMC implementation} \label{subsec:mcmc}

The parameters of light-curve models are obtained by Bayesian inference.

The different stochastic models are composed of a mean value $\bar{\phi}$, an AR term, and a white noise component $N(0,\sigma)$. For each stochastic model, the deterministic components are added to perform six different MCMC fits: pure noise, linear, sinusoidal, harmonic, linear + sinusoidal, linear + harmonic. 
This gives a total of 18 models computed. Finally, to account for observational uncertainties, each model includes a normal distribution $N(0,\epsilon_{obs})$ with standard deviation $\epsilon_{obs}$ corresponding to the measurement systematic $1\sigma$ errors in flux. 

The following list indicates the parameters and the mathematical description of each stochastic model and deterministic component to fit.

\textbf{Stochastic model:}
\begin{itemize}

    \item White Noise [$\bar{\phi}$, $\sigma$]:  $\phi(t_n) = \bar{\phi} + N(0, \sigma) \inlineeqnum$ 
   
    \item AR(1) [$\bar{\phi}$, $\sigma$, $\beta1$]:  $\phi(t_n) = \bar{\phi} + \beta_1 \phi(t_{n-1}) + N(0, \sigma) \inlineeqnum$ 
    
    \item AR(2) [$\bar{\phi}$, $\sigma$, $\beta1$, $\beta2$]:  $\phi(t_n) = \bar{\phi} + \beta_1 \phi(t_{n-1}) + \beta_2 \phi(t_{n-2}) + N(0, \sigma) \inlineeqnum$ 
\end{itemize}

\textbf{Deterministic component:}
\begin{itemize}

    \item  Linear [$C$]:  $ Ct_n \inlineeqnum$

    \item  Sinusoidal [$A_j$, $B_j$, $\omega_j$]:  $ \sum_{j} (A_j \cos(\omega_j t_n) + B_j \sin(\omega_j t_n)) \inlineeqnum$
    
    \item Harmonic [$A_j$, $B_j$, $A'_j$, $B'_j$, $\omega_j$]: $ \sum_{j} (A_j \cos(\omega_j t_n) + B_j \sin(\omega_j t_n) + A'_j \cos(2 \omega_j t_n) + B'_j \sin(2 \omega_j t_n)) \inlineeqnum$
    
\end{itemize}

There are between  2 (for a pure white noise model) and 11 or more parameters (for a AR(2) model with a linear term, a sinusoidal term and its harmonics). Taking the example of an AR(2) model with a single period and no linear term, the conditional probability of obtaining a lflux $\phi(t_n)$ is
\begin{equation}
P(\phi(t_n)| \phi(t_{n-1}), \phi(t_{n-2},\bar{\phi},\beta_1,\beta_1, C, \omega_i, A_i,B_i,\sigma.. ) = N(\bar{\phi} + \beta_1 \phi(t_{n-1}) + \beta_2 \phi(t_{n-2})+A \cos(\omega t_n) + B \sin(\omega t_n) )
\end{equation}    
The likelihood for the parameters $\bar{\phi},\beta_1,\beta_1, C, \omega_i, A_i,B_i,\sigma$ is \citep{Robert2007}
\begin{equation}
    L(\bar{\phi},\beta_1,\beta_1, C, \omega_i, A_i,B_i,\sigma..)=\frac{1}{\sigma^{\mathrm{ntot}}}\prod_0^\mathrm{ntot} \exp{-\frac{(\phi(t_n)-(\bar{\phi} + \beta_1 \phi(t_{n-1}) + \beta_2 \phi(t_{n-2})+A \cos(\omega t_n) + B \sin(\omega t_n) ))^2}{2\sigma_1^2}}
    \label{eq:lik}
\end{equation}
where $\mathrm{ntot}$ is the number of measurements and $\sigma_1^2 = \sigma^2+\mathrm{err}(t_n)^2$ takes into account the lflux measurement error at time $t_n.$  

In order to avoid working with numbers in different orders of magnitude, which may result in lower efficiency for the MCMC sampling, data are standardized by being re-scaled to their mean ($\bar{x}$) and std ($s_{x}$).:
\begin{equation}
\phi(t)_{st} = \frac{\phi(t) - \bar{\phi}(t)}{s_{\phi(t)}} \end{equation}
\begin{equation}
t_{st} = \frac{t - \bar{t}}{s_{t}}
\end{equation}

By Bayes theorem, the probability distribution of the parameters is the product of the likelihood (equation (\ref{eq:lik})) and 
priors on every parameter. 

The prior distribution of the MCMC parameters are selected to be as vague and noninformative as possible for the data sample analysed. This allows to minimize the influence and bias on the parameters posterior inference.

The standardization of the data also helps to set the scale for the prior distributions. Thus, the priors of the offset ($\bar{\phi}$), the amplitude terms ($A, B$), the slope ($C$) and the auto-regressive terms ($\beta_1, \beta_2$) were chosen as a normal distribution around 0 with a standard deviation of 2. 

For the period parameter, the posterior distribution is limited between a minimum and a maximum value. In the original scale, the lower value is set to 500 days to avoid the MCMC chains to be stuck in a possible artificial period\footnote{\url{https://fermi.gsfc.nasa.gov/ssc/data/analysis/LAT_caveats_temporal.html}} of 1 year and close values. The upper limit is set slightly above the half of the data time span ($\sim 2200$ days), so the period detection can be representative.
Now, the prior distribution is a normal centered in the middle of the space drawn for this parameter ($\sim 1350$ days), with a standard deviation of $\sim 800$ days. 

Finally, the white noise parameter ($\sigma$) is uniformly distributed between $[1\times 10^{-3}, 1\times 10^{3}]$. After the completion of the MCMC, the parameter outputs are transformed back to the original scale.

The models have been implemented in the R version of JAGS \citep{2012ascl.soft09002P}. JAGS is a Markov Chain Monte Carlo (MCMC) based on the Gibbs sampling algorithm. For each source, three independent chains were run, using 8000 iterations with a burning length of 4000 samples. The convergence of the chains was checked with the Gelman-Rubin diagnostic \citep{10.1214/ss/1177011136}. The output of the program 
is a set of posterior probability distributions, one for each 
parameter included in the model computed. Systematics 
of the output such as the prior dependence, the correlation between parameters and normality of residuals are described in Appendix.
Results from Table \ref{tab:mcmc} and \ref{tab:slices} quote the mean values of the posterior distributions and 95\% credible intervals around the mean. 

As explained in the next section, the fits also allow performing deviance comparisons between periodic and pure noise non-periodic models.

\subsubsection{Model Selection through Information Theory} \label{subsec:model}

Information theory is introduced in model selection problems as a form of quantitative explanation of the best model’s goodness of fit. \citet{1100705} proposed a way to estimate divergence based on the maximized empirical log-likelihood estimator (MLE), the Akaike Information Criterion (AIC). AIC is used as a measure of the information lost when the fitted model is used to approximate the process that generates the empirical data:
\begin{equation}
AIC = -2\log\mathcal{L}(\theta|y) + 2K
\label{eq:aic}
\end{equation}
where $\log\mathcal{L}(\theta|y)$ is the log likelihood of the model given the data $y$, K is the number of model's parameters (defined as $\theta$) and operates as a form of penalty to model complexity. Thus, AIC is an effective tool for selecting a simple model which describes and infers empirical data, avoiding both overfitting and underfitting.
In our MCMC pipeline, the assessment is done using:
\begin{equation}
    AIC = D(\theta|y)_{min} + 2K
\label{eq:aic_mcmc}
\end{equation}
where $D(\theta|y)_{min} = -2\log\mathcal{L}(\theta|y)$ is the minimum deviance of the MCMC posterior sample.

For each light-curve, within all the possible MCMC implementations, the one with the minimum AIC is selected. Now, for model statistical assessment, AIC does not carry much information as it is in a relative scale and it is dependent on sample size. What matters in model assessment, though, is $\Delta AIC$, the difference between AIC values over multiple nested models. Given a full model $F$ and a reduce model $R$:
\begin{equation}
    \Delta AIC = AIC_R - AIC_F = -2\log(\mathcal{L}(\theta_0)/\mathcal{L}(\theta)) - 2k = \Lambda - 2k
\label{eq:dAIC}
\end{equation}
where $\Lambda$ is the likelihood ratio test statistic, $\mathcal{L}(\theta_0)$ and $\mathcal{L}(\theta)$ are the MLE under the null ($R$) and alternative ($F$) hypothesis, respectively, and $k$ is the difference of parameters between models.

From this definition, a $p_{value}$ can be computed, which shows the probability of obtaining the value $\Lambda$ under the null hypothesis conditions.
As stated in \citet{EfronHastie}, chapter 13 (see also \citet{Murtaugh}), the relationship between $p_{value}$ and $\Delta AIC$ can be drawn as:
\begin{equation}
    p_{value} = Pr(\chi^2_k > \Lambda) = Pr(\chi^2_k > \Delta AIC + 2k)
    \label{eq:pvalue}
\end{equation}
where $\Lambda = \Delta AIC + 2k$ follows a $\chi^2$ distribution with $k$ degrees of freedom. 
In the following we therefore use this relation to assess a $p_{value}$ value when comparing nested models.

\subsection{Spectral Method} \label{subsec:agatha}

In this paper, potential periods of light-curves are searched by a time-domain, MCMC based, method, introduced in previous section. To validate our results, we found useful to compare the periods found with those obtained with a different, spectral based, method.
The public domain Agatha program \citep{2017MNRAS.470.4794F} calculates Lomb-Scargle periodograms 
on astronomical time-series and evaluates the significance of the periodic components found. Four different variants of the Lomb-Scargle algorithm are implemented and were studied with simulated light-curves. The Bayes Factor Periodogram (BFP) was found to be efficient at finding periods in noisy light curves with trend. To asses the significance of the computed periodograms, the logarithm of the Bayes Factor (lnBF) is evaluated. lnBF is related to the maximum likelihood ratio of the periodic and the noise model and can be seen as the significance of the given period.
Agatha provides also a "moving periodogram" (periodogram calculated in different time windows) which is a useful tool for finding long term changes in periods, studied in section \ref{subsec:slices}. The authors of Agatha recommend to use their program in combination with a MCMC search to refine the results. As explained in the previous section, the priors of our MCMC search are not based on Agatha results, which are only used as cross-checks.

\section{Fermi Data} \label{sec:data}

This section is dedicated to explain the selection and analysis of the sample of AGNs light curves from the Fermi-LAT data.

\subsection{Object Selection} \label{subsec:selection}

The sources selected in this work compose a subsample of the AGN population of the 4FGL,  the Fermi LAT 8-year Source Catalog \citep{Abdollahi_2020}.  The 4FGL represents a daily full sky survey in the 50 MeV - 1 TeV energy range.  From the total of 5064 sources in the catalog, 3207 are tagged as AGNs of which 3137 are blazars, 42  are radio galaxies and 28 are other kind of AGNs.
Our AGN selection was motivated by previous studies on period detection in gamma-ray literature. Information about the sources' properties and periodicity literature results is shown in Table~\ref{tab:objects}.
\begin{deluxetable*}{ccccc}
\tablenum{1}
\tablecaption{List of Fermi-LAT AGN sample with 4FGL and Common Name, source type, detected period in literature and reference.
\label{tab:objects}}
\tablewidth{0pt}
\tablehead{
\colhead{4FGL Name} & \colhead{Common name} &  \colhead{Type} & 
\colhead{Period (days)} & \colhead{Reference}
}
\startdata
J0043.8+3425 & GB6 J0043+3426 & FSRQ & 657 & 3\\
J0102.8+5825 & TXS 0059+581 & FSRQ & 767 & 3\\
J0158.5+0133 & 4C +01.28 & BL Lac & 445 & 4\\
J0210.7-5101 & PKS 0208-512 & FSRQ & 949 & 3\\
J0211.2+1051 & CGRaBS J0211+1051 & BL Lac & 621 & 3\\
J0252.8-2218 & PKS 0250-225 & FSRQ & 438 & 3\\
J0303.4-2407 & PKS 0301-243 & BL Lac & 730, 766$\pm$109 & 3, 9\\
J0428.6-3756 & QSO B0426-380 & BL Lac & 1241, 1223$\pm$248 & 3, 8\\
J0449.4-4350 & PKS 0447-439 & BL Lac & 913 & 3\\
J0457.0-2324 & QSO J0457-2324 & FSRQ & 949 & 3\\
J0501.2-0158 & PKS 0458-02 & FSRQ & 621 & 3\\
J0521.7+2112 & RX J0521.7+2112 & BL Lac & 1022 & 3\\
J0721.9+7120 & PKS 0716+71 & BL Lac & 1022, 346 & 3, 4\\
J0808.2-0751 & QSO B0805-077 & FSRQ & 658 & 4\\
J0811.4+0146 & QSO B0808+019 & BL Lac & 1570 & 3\\
J0818.2+4222 & QSO B0814+42 & BL Lac & 803 & 3\\
J1146.9+3958 & B2 1144+40 & FSRQ & 1205 & 3\\
J1248.3+5820 & QSO B1246+586 & BL Lac & 803 & 3\\
J1303.0+2434 & MG2 J130304+2434 & BL Lac & 730 & 3\\
J1454.4+5124 & TXS 1452+516 & BL Lac & 767 & 3\\
J1555.7+1111 & PG 1553+113 & BL Lac & 790, 803, 798, 780$\pm$63, 803 & 2, 3, 4, 5, 6 \\
J1649.4+5235 & 87GB 164812.2+524023 & BL Lac & 986 & 3\\
J1903.2+5540 & 1RXS J190313.1+554035 & BL Lac & 1387 & 3\\
J2056.2-4714 & PMN J2056-4714 & FSRQ & 620, 637 & 3, 4\\
J2158.8-3013 & PKS 2155-304 & BL Lac & 685, 610, 621, 644, 620$\pm$41, 635$\pm$47 & 1, 2, 3, 4, 5, 7 \\
J2202.7+4216 & BL Lac & BL Lac & 698, 680$\pm$35 & 4, 5 \\
J2258.1-2759 & VSOP J2258-2758 & FSRQ & 475 & 3
\enddata
\vspace{1ex}
{\raggedright \textbf{References:}(1) \citet{2019MNRAS.484..749C} (2) \citet{2020ApJ...895..122C} (3) \citet{2020ApJ...896..134P} (4) \citet{2017MNRAS.471.3036P} (5) \citet{2018AA...615A.118S} (6) \citet{2018ApJ...854...11T} (7) \citet{2017ApJ...835..260Z} (8) \citet{2017ApJ...842...10Z} (9) \citet{2017ApJ...845...82Z} \par}
\end{deluxetable*}

\subsection{Data Analysis} \label{subsec:analysis}

The analysis of each source was performed using Enrico, a community-developed Python package to conduct Fermi-LAT analysis \citep{2013ICRC...33.2784S}, which consist in a simplified full analysis chain based on the FermiTools\footnote{\url{https://github.com/fermi-lat/Fermitools-conda}}. The version of the Fermitools is 2.0.8 via the conda repository. 

The data are obtained from Fermi-LAT Data Server\footnote{\url{https://fermi.gsfc.nasa.gov/cgi-bin/ssc/LAT/LATDataQuery.cgi}} introducing Astroquery \citep{2019AJ....157...98G} in our pipeline. 
More than 12 years of Pass 8 LAT data \citep{2013arXiv1303.3514A} are used with \texttt{evclass=128} and \texttt{evtype=3} for photon-like events in point sources analysis. Events with zenith angle greater than $100^{\circ}$ are rejected to reduce the gamma-ray contamination of the Earth limb. Good time intervals with high quality data are selected using \texttt{(DATA\_QUAL$>$0)\&\&(LAT\_CONFIG==1)}.
On sky model generation, the background emission is modeled adopting the Galactic diffuse emission file \texttt{gll\_iem\_v07.fits} and the extragalactic isotropic diffuse emission file \texttt{iso\_P8R3\_SOURCE\_V2\_v1.txt}.

The analysis software computes a binned likelihood analysis to find the best fit model parameters and the light-curves are obtained by running the entire chain into time bins.
The light-curves between 1 GeV and 300 GeV were computed in 145 time bins from the start of the mission (239557418 MET) until the end of March 2021 (638767517 MET), using an ROI of 10 degrees. Within each time bin, the pipeline generates a light curve point unless the TS is under 9, in which case an upper limit is derived.

Finally, as justified in Section \ref{sec:methods}, the logarithm of the flux is applied. For the study on periodicity, only light curves without important flux gaps are considered. Thus, the final AGN sample is limited to the sources included in the following section.

\section{Results} \label{sec:results}

\subsection{Full Light-Curve}\label{sec:fulllc}
The light-curves obtained after the analysis chain explained in \ref{subsec:analysis} are analysed following the MCMC procedures in \ref{subsec:mcmc}. The results are shown in Table \ref{tab:mcmc}. 
\begin{deluxetable*}{ccccccc|cc}
\tablenum{2}
\tablecaption{
MCMC fit results and Agatha cross-check comparison of the AGN Fermi-LAT sample.  For each source, the list indicates: the best model in terms of AIC; the AIC value; the period mean and standard deviation in days; the period 95\% Highest Density Intervals in days; the $\Delta$AIC between the periodic and the noise model; the p-value computed from $\Delta$AIC; the Agatha period mean and standard deviation; the Agatha lnBF. * indicates specific prior assumption and **~indicates inferior posterior convergence for period (see Appendix \ref{ap:prior}).
The results are sorted by MCMC period detection significance. \label{tab:mcmc}}
\tablewidth{0pt}
\tablehead{
\colhead{4FGL Name} & \colhead{Best Model} & \colhead{AIC} & \colhead{Period} & \colhead{Period HDI$_{95\%}$} & \colhead{$\Delta$AIC} & \colhead{$p_{value}$} & 
\colhead{Period\textsubscript{AGATHA}} &
\colhead{lnBF}
}
\startdata
J1555.7+1111 & AR(1) + lin + sin & 298.03 & 774 $\pm$ 10 & 755 - 793 & 28.98 & 1.2$\times 10^{-7}$ & 771 $\pm$ 29 & 10.9 \\
J2158.8-3013 & AR(1) + lin + sin & 340.07 & 614 $\pm$ 16 & 589 - 642 & 9.85 & 1.2$\times 10^{-3}$ & 615 $\pm$ 26 & 6.3 \\
J1903.2+5540 & AR(1) + lin + sin & 381.92 & 1120 $\pm$ 95 & 1040 - 1230 & 8.7 & 2.1$\times 10^{-3}$ & 1163 $\pm$ 55 & 0.6 \\
J0303.4-2407 & AR(1) + lin + harm & 323.83 & 821 $\pm$ 40 & 761 - 870 & 5.58 & 8.2$\times 10^{-3}$ & 773 $\pm$ 26 & -1.2 \\
J0521.7+2112 & AR(2) + sin & 309.1 & 1136 $\pm$ 128 & 990 - 1280 & 5.52 & 9.2$\times 10^{-3}$ & 1139 $\pm$ 73 & 11 \\
J1248.3+5820* & AR(2) + sin & 383.73 & 2048 $\pm$ 169 &  1800 - 2350 &  4.29 &  1.6$\times 10^{-2}$ & 2039 $\pm$ 133 & 3 \\
J0211.2+1051 & AR(1) + harm & 301.31 &  1398 $\pm$ 122 &  1190 - 1630 & 3.62 & 1.8$\times 10^{-2}$ & 1446 $\pm$ 59 & 3 \\
J0449.4-4350** & AR(1) + lin + sin &  296.29 &  746 $\pm$ 229 &  505 - 1030 &  3.6 &  2.2$\times 10^{-2}$ & 669 $\pm$ 14 & 7.2 \\
J0457.0-2324* & AR(1) + sin & 293.61 &  1300 $\pm$ 153 &   975 - 1590 & 2.11 &  4.4$\times 10^{-2}$ & 1330 $\pm$ 59 & 7.9 \\
J2202.7+4216* & AR(1) + lin + sin & 261.19 & 1799 $\pm$ 219 & 1430 - 2250 & 2.01 & 4.6$\times 10^{-2}$ & 1763 $\pm$ 89 & 0.1 \\
J0721.9+7120** & AR(1) + sin &  321.08 &  987 $\pm$ 220 &  574 - 1520 & 1.94 &  4.7 $\times 10^{-2}$ & 1011 $\pm$ 96 & 7.3 \\
J0818.2+4222** & AR(2) + sin & 360.61 & 955 $\pm$ 356 & 501 - 1790 & 1.68 & 5.3$\times 10^{-2}$ & 1333 $\pm$ 21 & 2.1 \\
J0428.6-3756* & AR(1) + lin + sin & 288.65 &  1310 $\pm$ 175 & 889 - 1650 & 0.94 & 7.4$\times 10^{-2}$ & 1262 $\pm$ 114 & 13.2 \\
J0210.7-5101** & AR(2) + lin + sin & 210.93 & 1080 $\pm$ 351 & 502 - 1640 & 0.68 & 8.3$\times 10^{-2}$ & 1025 $\pm$ 13 & 3.9 
\enddata
\end{deluxetable*}

All AGNs analyzed present a correlated colored noise, depicted from the auto-regressive AR(N) terms. The inclusion of these components over white noise is fundamental in terms of significance, regarding the model selection procedure presented in Section \ref{subsec:model}. The $\Delta AIC_{wc} = AIC_{white} - AIC_{colored}$ are between 10 and 60 for all sources, which is more than sufficient to reject a white noise model over a colored noise one. 
Nonetheless, it is important to remark that using an auto-regressive noise model decreases the significance of periodic signals compared to using a white noise models.
This means that, in most sources analyzed, $\Delta AIC_w = AIC_{white \; noise} - AIC_{white\; periodic}$ is greater than $\Delta AIC_c = AIC_{colored \; noise} - AIC_{colored \; periodic}$.

\begin{figure}[t!]
\centering
\includegraphics[width = 0.49\textwidth]{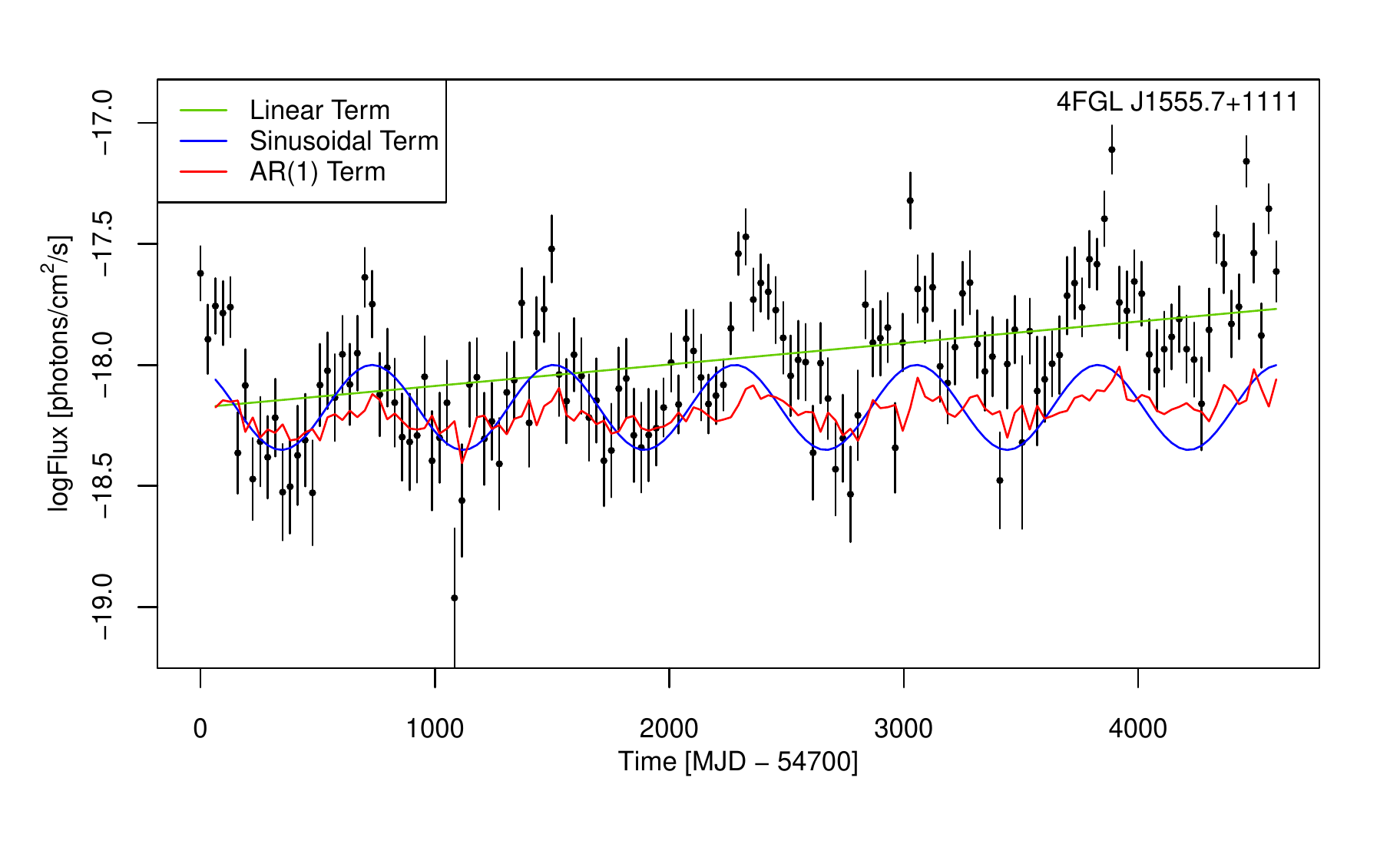}
\includegraphics[width = 0.49\textwidth]{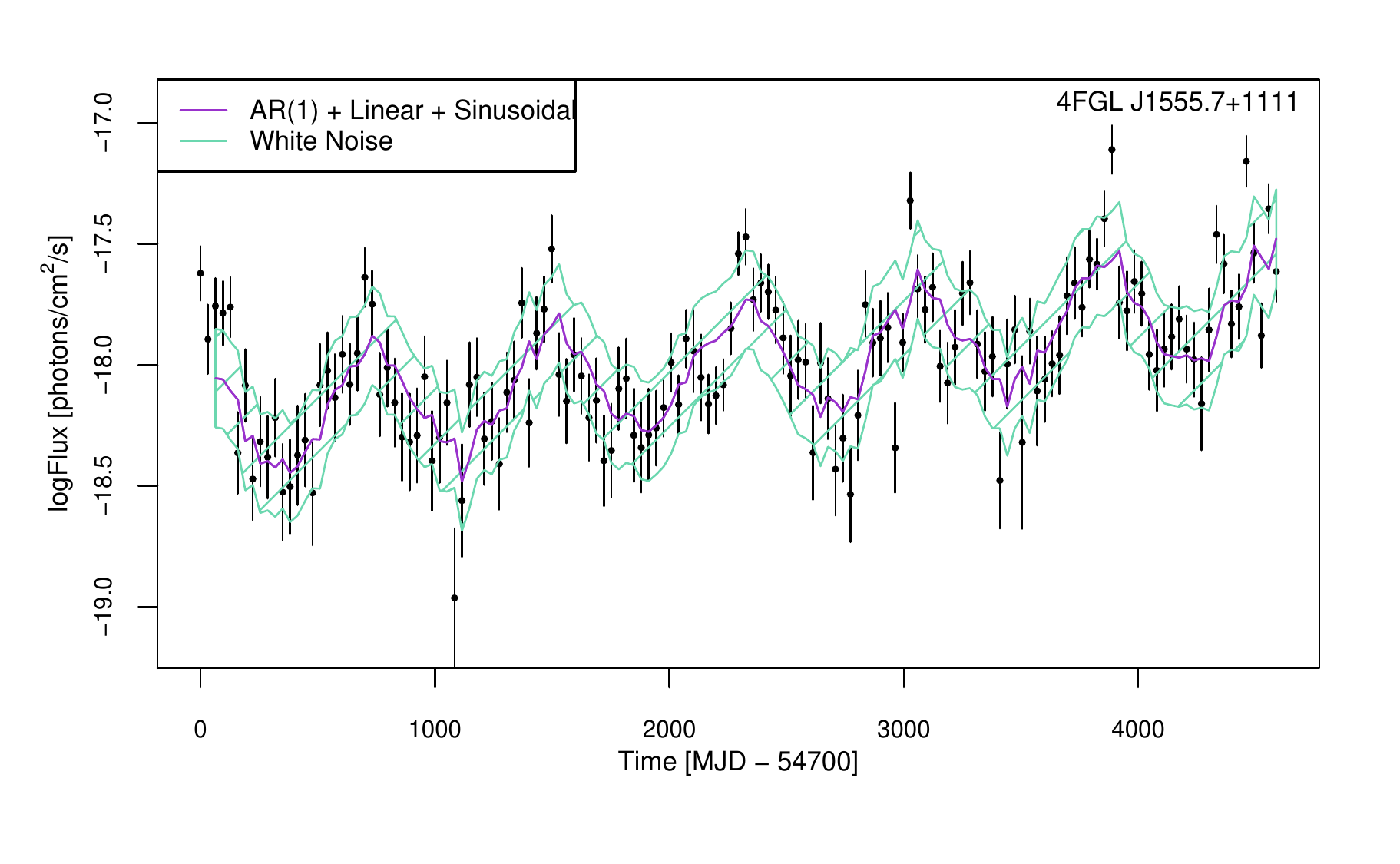}
\includegraphics[width = 0.49\textwidth]{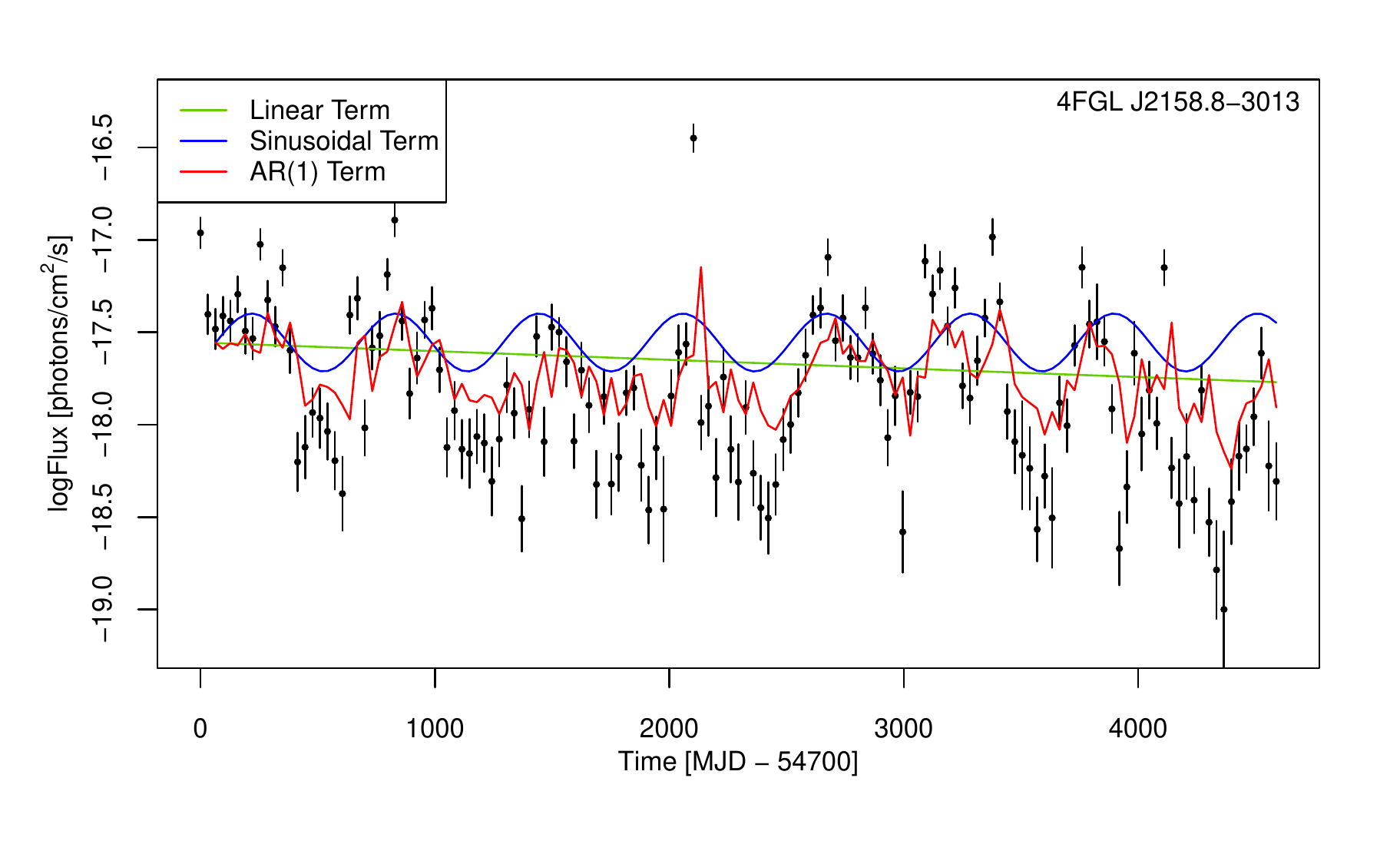}
\includegraphics[width = 0.49\textwidth]{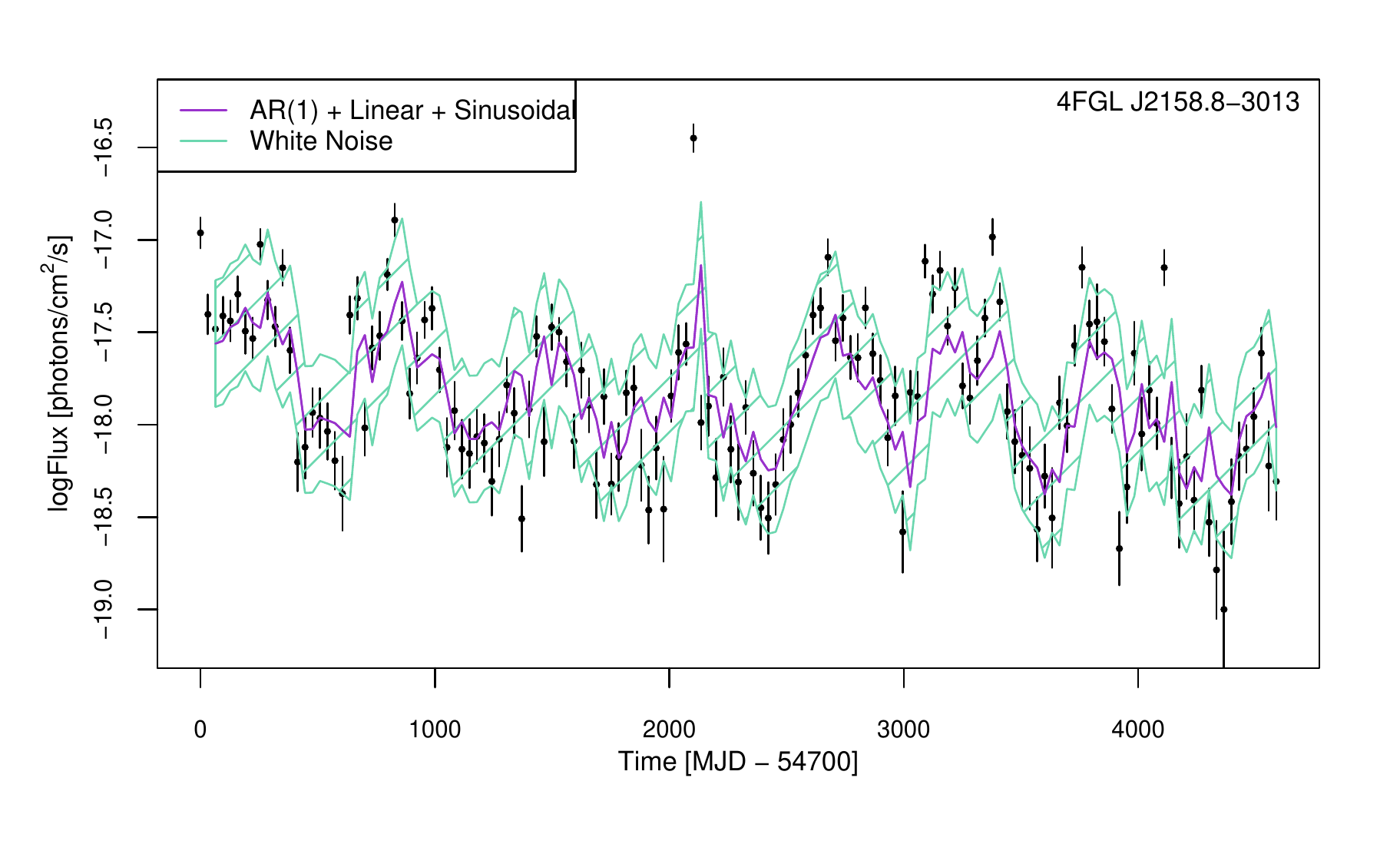}
\includegraphics[width = 0.49\textwidth]{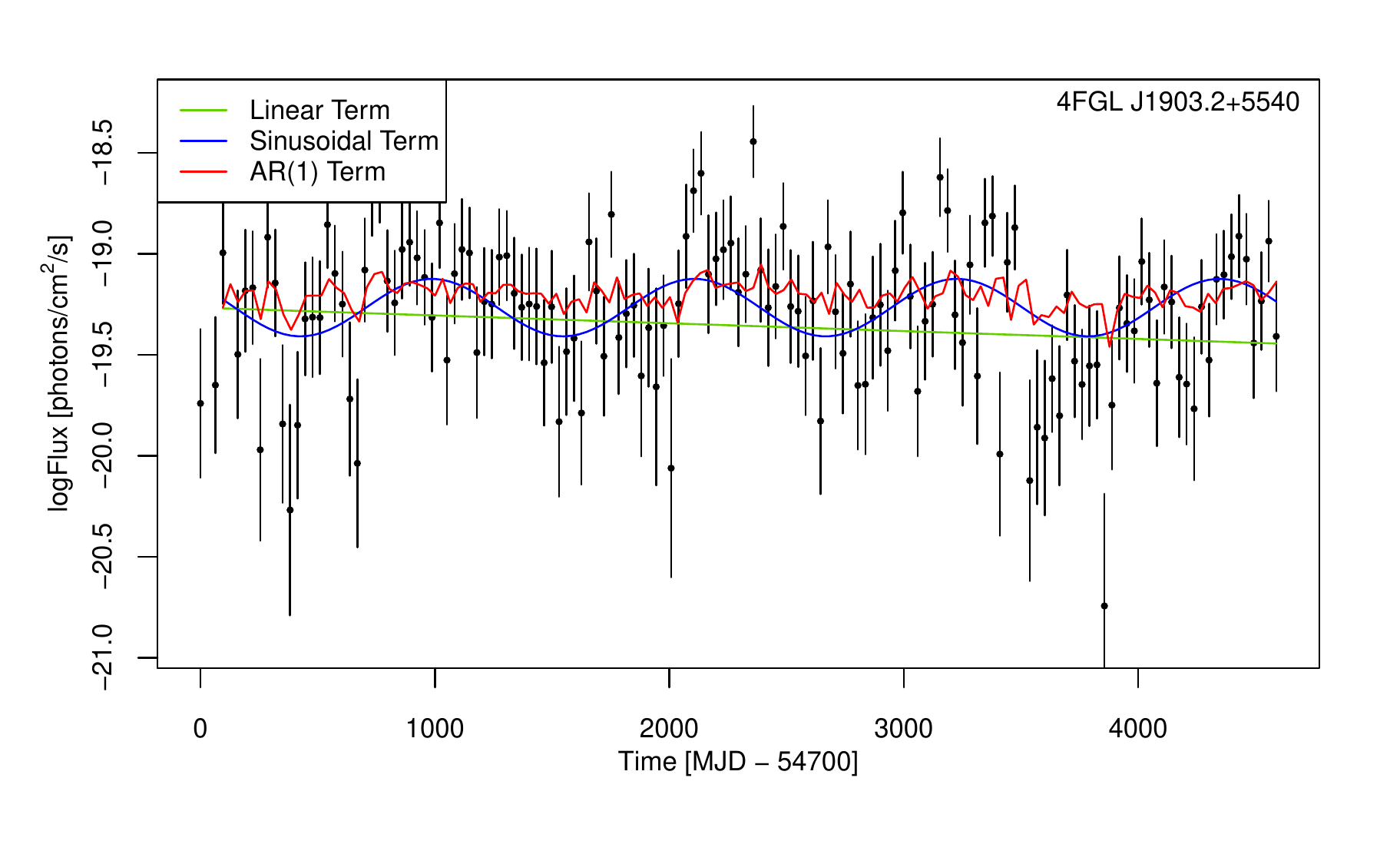}
\includegraphics[width = 0.49\textwidth]{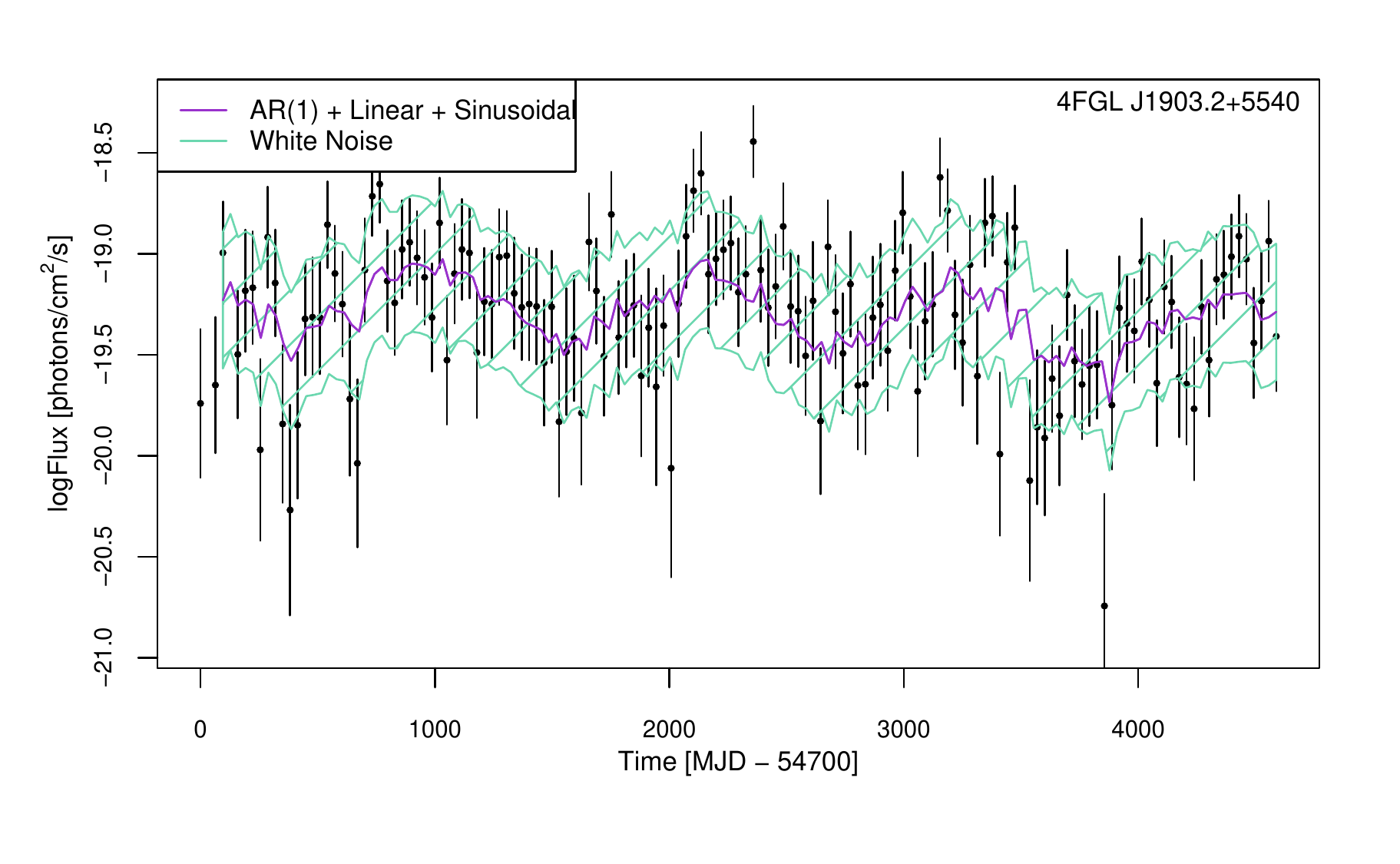}
\caption{MCMC fit for the three sources with most significant periodic signals. [\textit{Left Panel}] Separated fitted components of the model. Green line indicates the linear trend. Blue line indicates the sinusoidal term. Red line indicates the stochastic term [\textit{Right Panel}] General fit and White Noise $\sigma$ component.   \label{fig:mcmcfits}}
\end{figure}
For all sources, a periodic model is preferred and the periodic signal is assessed through the computation of the $p_{value}$ by its relationship with $\Delta AIC$ (eq.\ref{eq:pvalue}). 

The most significant periodic signal is that for PG~1553+113, with a $p_{value}$ of $1.2 \times 10^{-7}$ denoting a very strong evidence of the source's periodic behaviour. The period found of 774 days($\sim 2.1$  years) is compatible with previous periodicity studies on the source.
Besides, the light-curve shows a clear linear trend $C= 9\times10^{-5}$ and a small noise/auto-regressive term $\beta_1 = 0.22$.
The inclusion of a linear trend in the fit of the source's behaviour is also fundamental in terms of significance of periodicity. In the fit without a linear trend, the same period is found with a much lower $p_{value}'= 3.4\times10^{-4}$ and a much higher noise term $\beta_1' = 0.49 $. 

The next sources with the most significant periodic signal are PKS~2155-304 and TXS~1902+556, with $p_{values}$ below 0.005. For TXS~1902+556, a period of 1120 days ($\sim 3.1$ years) is identified. This value is not in agreement with previous literature periodic analysis, \cite{2020ApJ...896..134P}, where a period of $\sim 3.8$ years is found in the low significance level ($>2.5 \sigma$). On the other hand, for PKS~2155-304, the fitted periodic signal 614 days ($\sim 1.7$ years) is compatible with literature. Both sources show also a linear trend and an auto-regressive noise behaviour.
The aforementioned descriptions for the three most significant periodic source's MCMC fits are shown in Figure~\ref{fig:mcmcfits}.

\begin{figure}[t!]
\centering
\includegraphics[width = 0.49\textwidth]{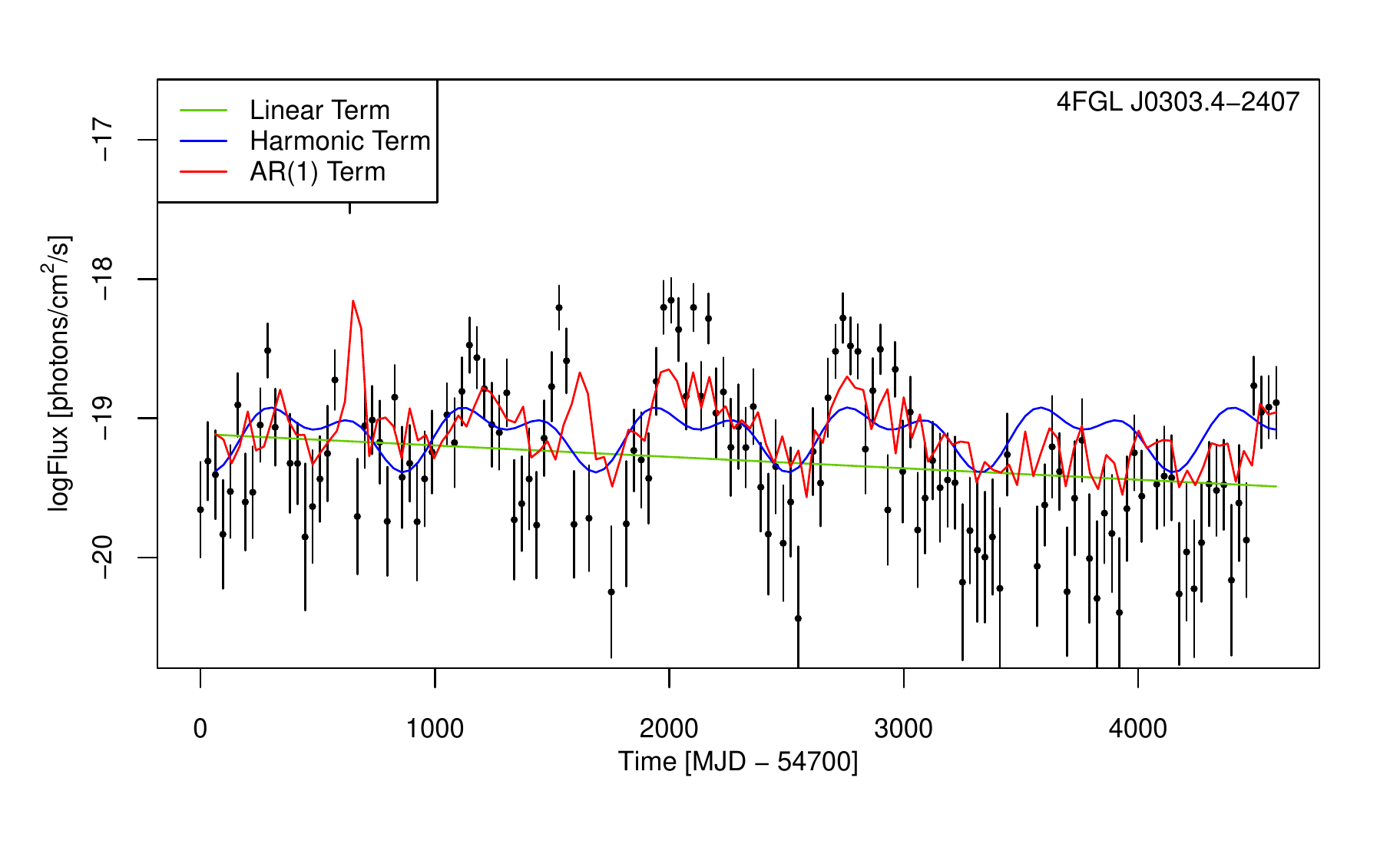}
\includegraphics[width = 0.49\textwidth]{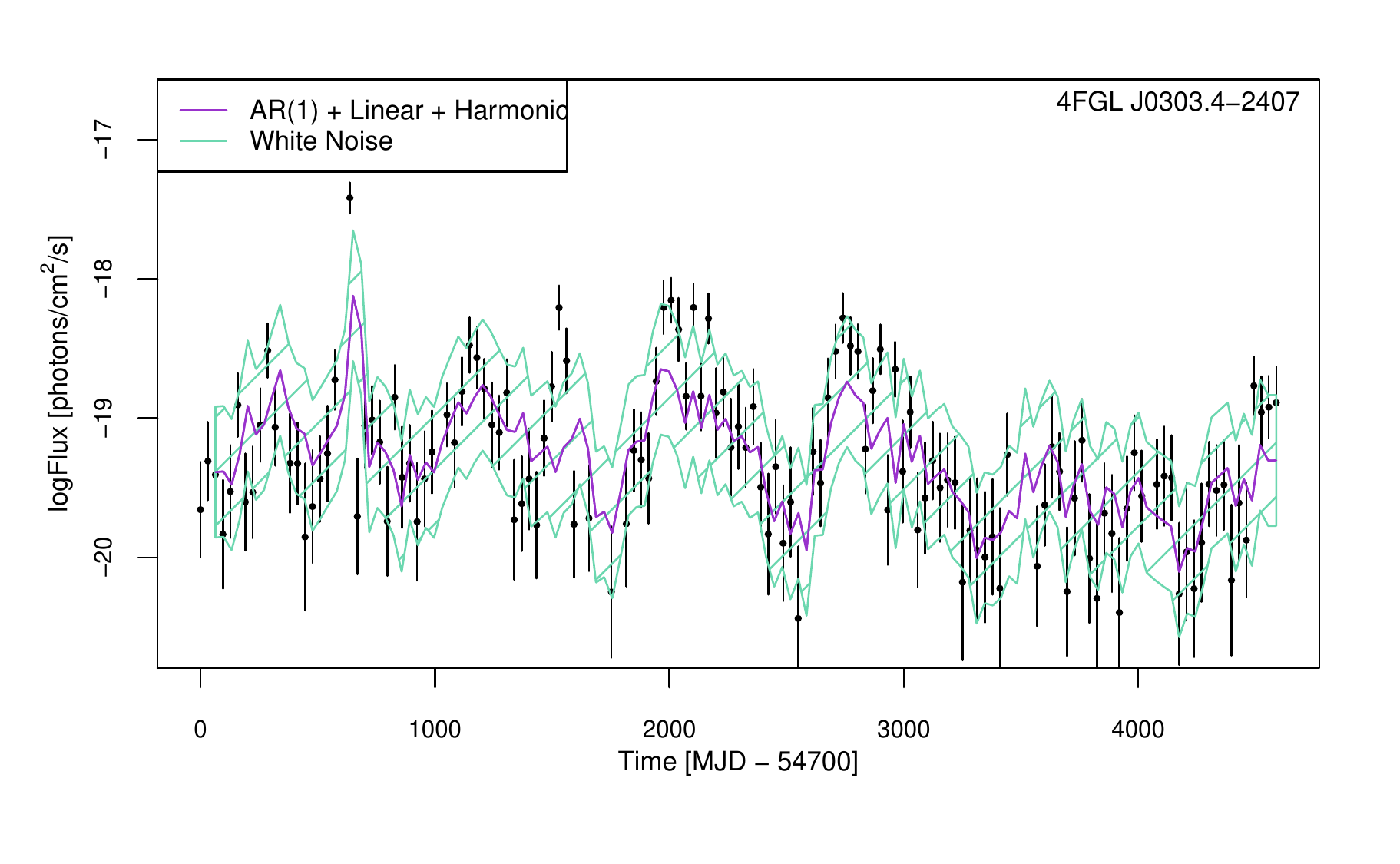}
\caption{MCMC fit for PKS~0301-243. [\textit{Left Panel}] Separated fitted components of the model. Green line indicates the linear trend. Blue line indicates the sinusoidal harmonic term. Red line indicates the stochastic term [\textit{Right Panel}] General fit and White Noise $\sigma$ component.   \label{fig:mcmcfits_harm}}
\end{figure}

Two sources, PKS~0301-243 and RX~J0521.7+2112, are below a $p_{value}$ of 0.01, with a period of 821 days ($\sim 2.3$ years) and 1136 days ($\sim 3.1$ years), respectively. Both values are higher and not compatible with those on literature, where periods of $\sim 2$  and $\sim 2.8$ years are found in the high significance level ($>3 \sigma$), respectively.  
RX~J0521.7+2112 does not show a linear trend and it is dominated by an auto-regressive component of second order with $\beta_1 = 0.38$ and $\beta_2 = 0.24$.
PKS~0301-243 shows a special sinusoidal behaviour, a principal periodic component of 821 days ($\sim 2.3$ years) with a second harmonic. The result on this source is shown in Figure \ref{fig:mcmcfits_harm}. For CGRaBS~J0211+1051 an harmonic oscillation of 1398 days ($\sim 3.8$ years) is also found with lower significance.
The appearance of an harmonic component in the sources light-curve will be discussed in section \ref{sec:discussion}.

The periodicity significance of the remaining AGNs in Table \ref{tab:mcmc} is low ($p_{value} >  0.01$) and the MCMC performance is not as good as for the high significance cases. As a result, the standard deviation of the period parameter increases ($> 150$ days) as well as the Highest Desnity Interval (HDI), which, for some of the sources, is stuck either in the lower or the upper posterior limit. Also, the colored noise terms are much higher, being all above 0.5.
For the sources marked with *, the period values are taken from a good posterior distribution  obtained by the use of specific priors different from the general prior in Section \ref{subsec:mcmc}. This is described in Appendix~\ref{ap:prior}. For the sources marked with **, the MCMC chains convergence for the period parameter is poorer, resulting in posterior distributions with shapes different from a expected symmetric Gaussian. An example is also shown in Appendix~\ref{ap:prior}. 

\begin{figure}[t!]
\centering
\includegraphics[width = 0.32\textwidth]{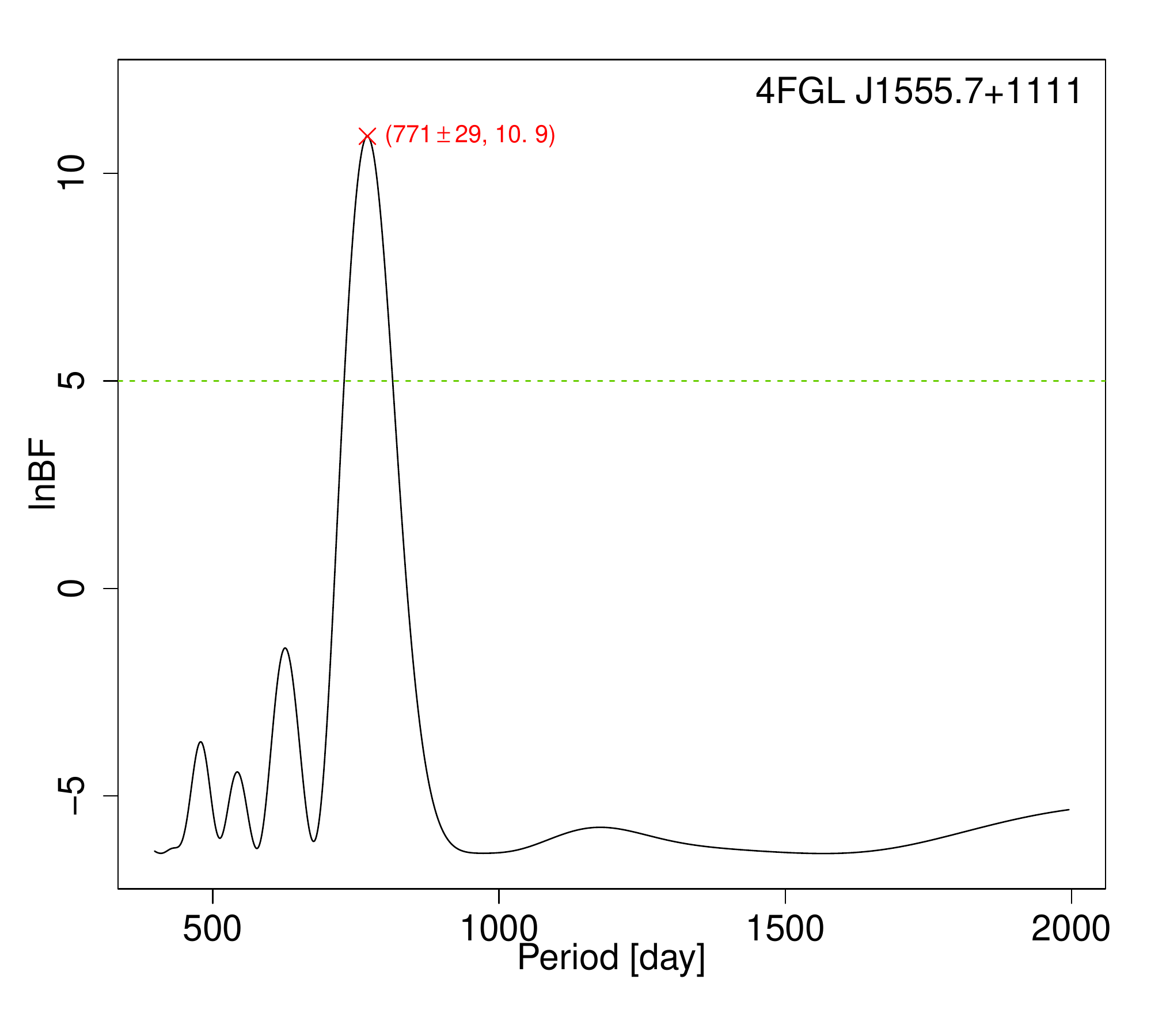}
\includegraphics[width = 0.32\textwidth]{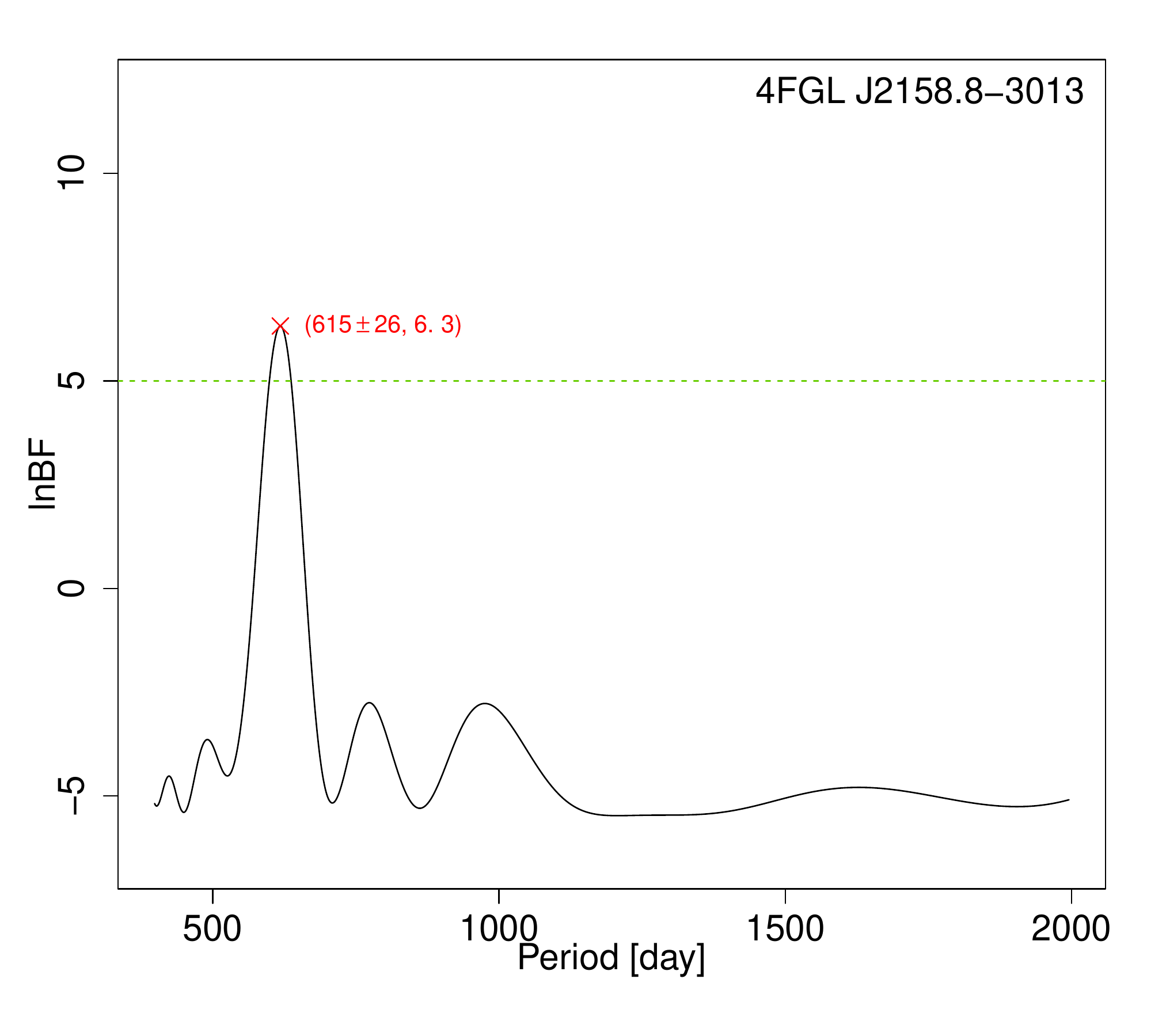}
\includegraphics[width = 0.32\textwidth]{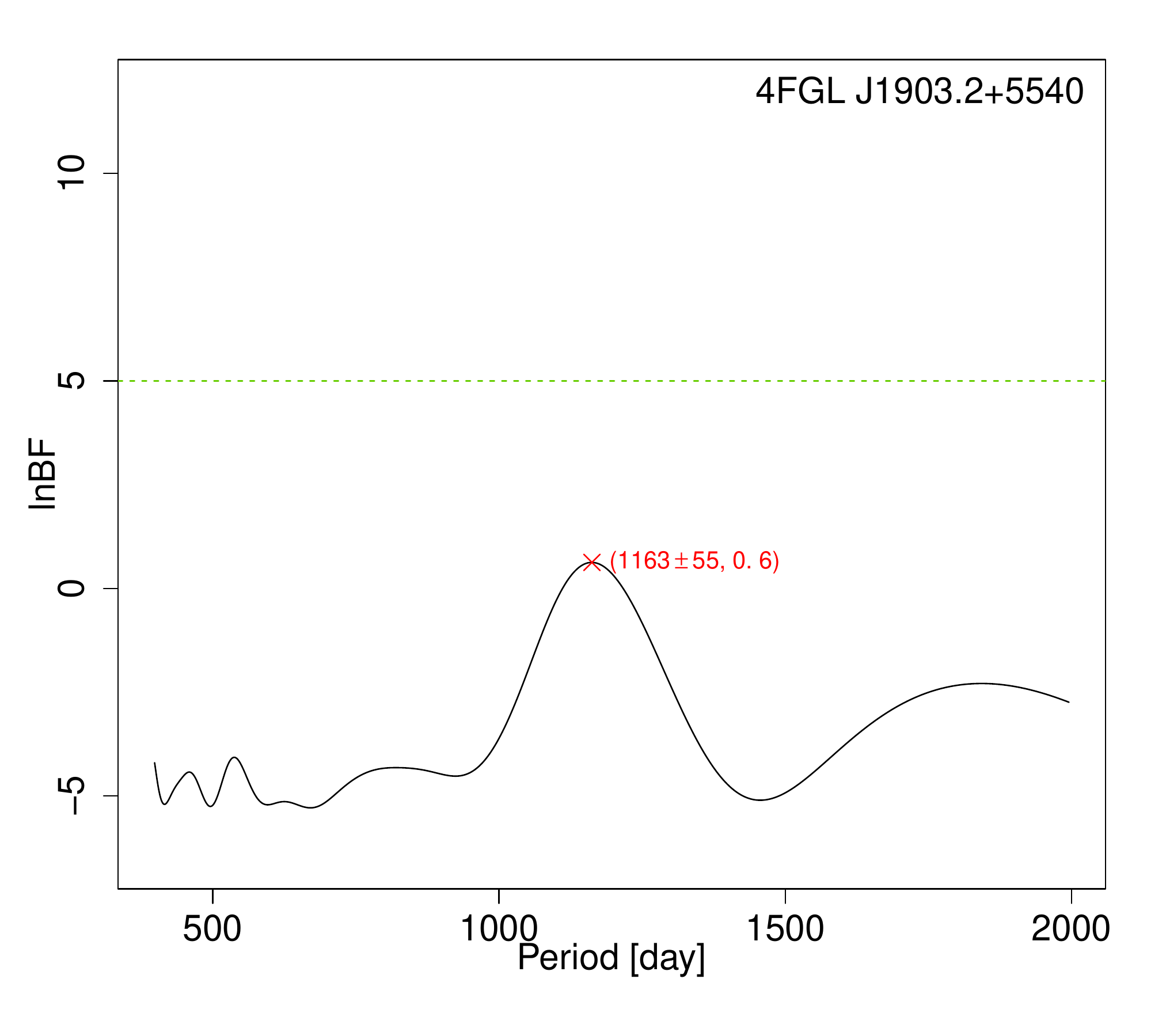}
\caption{Agatha results for the three sources with most significant MCMC sinusoidal signals. 
The green dashed line indicates the 5 lnBF significance \textbf{as prescribed by \citet{2017MNRAS.470.4794F}}. \label{fig:agatha}}
\end{figure}

After the completion of the Time-Series analysis through the MCMC fits, a spectral analysis, presented in Section \ref{subsec:agatha}, is performed as a cross-check. 
For each source, the BFP is performed using  the same AR noise model as the one retrieved in the MCMC fit. The results of the Agatha analysis can be found in the right part of Table \ref{tab:mcmc}. As an example, BFPs of the sources represented in Figure \ref{fig:mcmcfits} are shown in Figure \ref{fig:agatha}.

Of the 14 periodic light-curves analyzed, the Agatha periods for all sources but for 4FGL~J0818.2+4222 and 4FGL~J0303.4-2407 are compatible with the MCMC results. Nonetheless, the two most significant signals of 4FGL~J0818.2+4222 are 1337 days and 805 days, with very close lnBF values of 2.1 and 1.6 respectively. The second most significant signal at 805 days is compatible with the MCMC result at 867 days.
For 4FGL~J0303.4-2407, a lnBF=-1.2 means that a noise model is favored over the periodic model at this given period. This might be due to the presence of an harmonic signal in the MCMC fit, which is not reproducible with Agatha.
4FGL~J1903.2+5540 and 4FGL~J2202.7+421 have a lnBF=0.6 and lnBF=0.1 respectively, meaning that the periodic model is poorly favored over the noise model. 
All the remaining sources present compatible periods with an important level of significance, i.e., lnBF$\geq3$.

As will be discussed in Section \ref{sec:discussion}, the time stability of the period and the variation of the amplitude of the periodic modulation as a function of the period are two important clues of the physical mechanism causing the oscillations. The local period and amplitudes can be obtained by building a spectrogram with time windows of the light-curve.

\subsection{Time Windows} \label{subsec:slices}

\begin{figure}[t!]
\centering
\includegraphics[width = 0.45\textwidth]{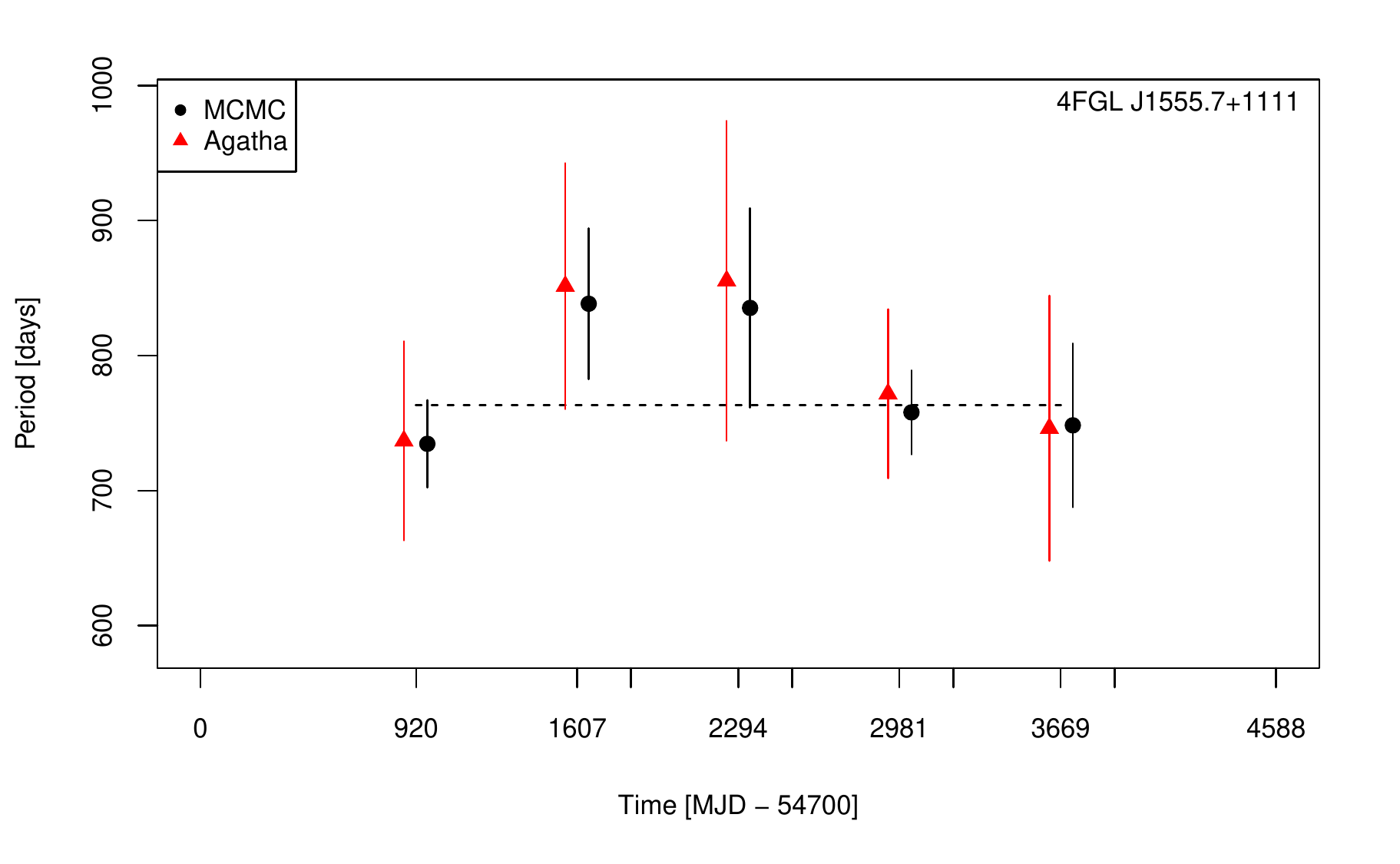}
\includegraphics[width = 0.45\textwidth]{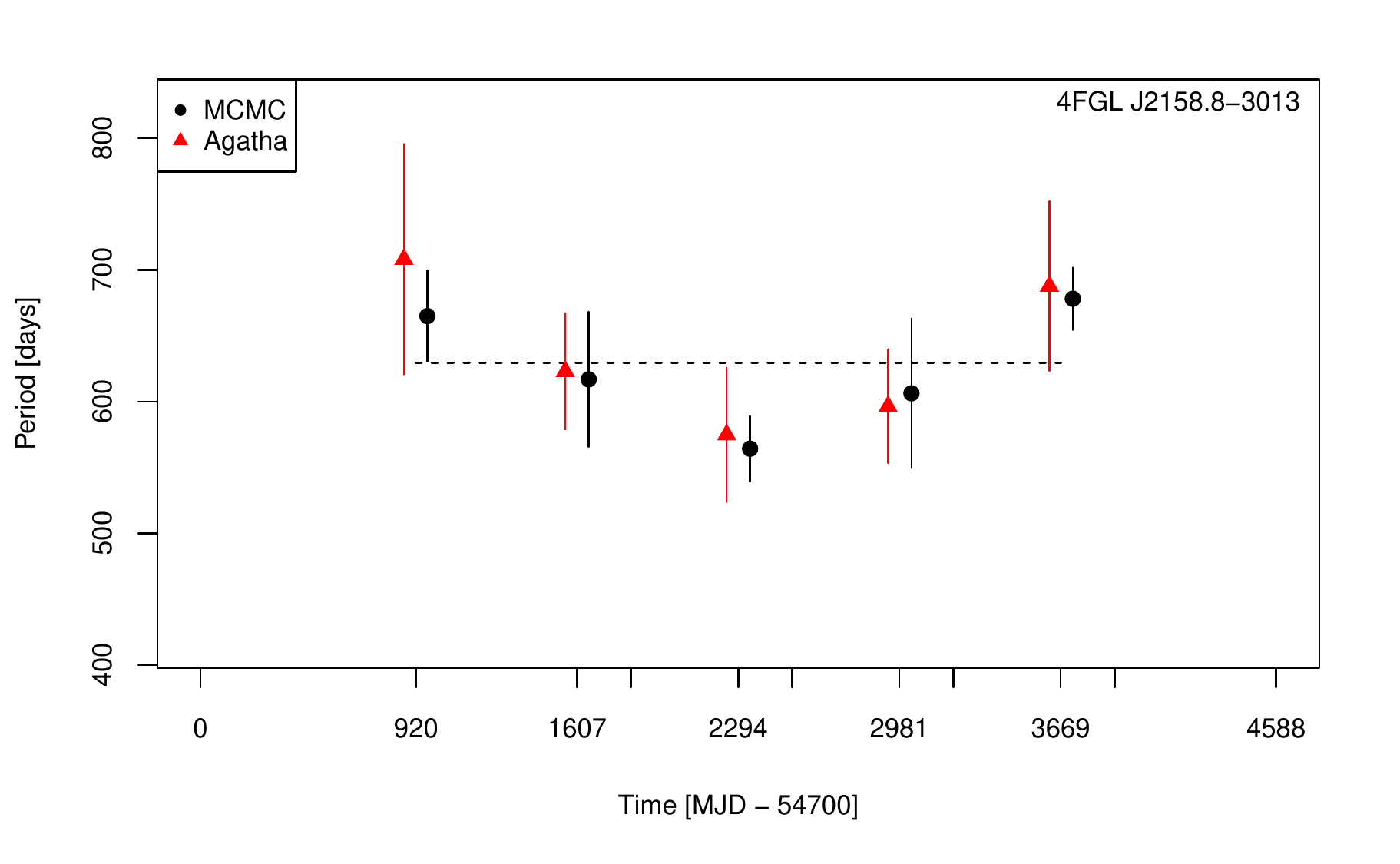}
\caption{MCMC and Agatha period mean and standard deviation for each time window. Horizontal dashed line indicates the result of the constant fit value.
 \label{fig:slice_p}}
\end{figure}

\begin{figure}[t!]
\centering
\includegraphics[width = 0.45\textwidth]{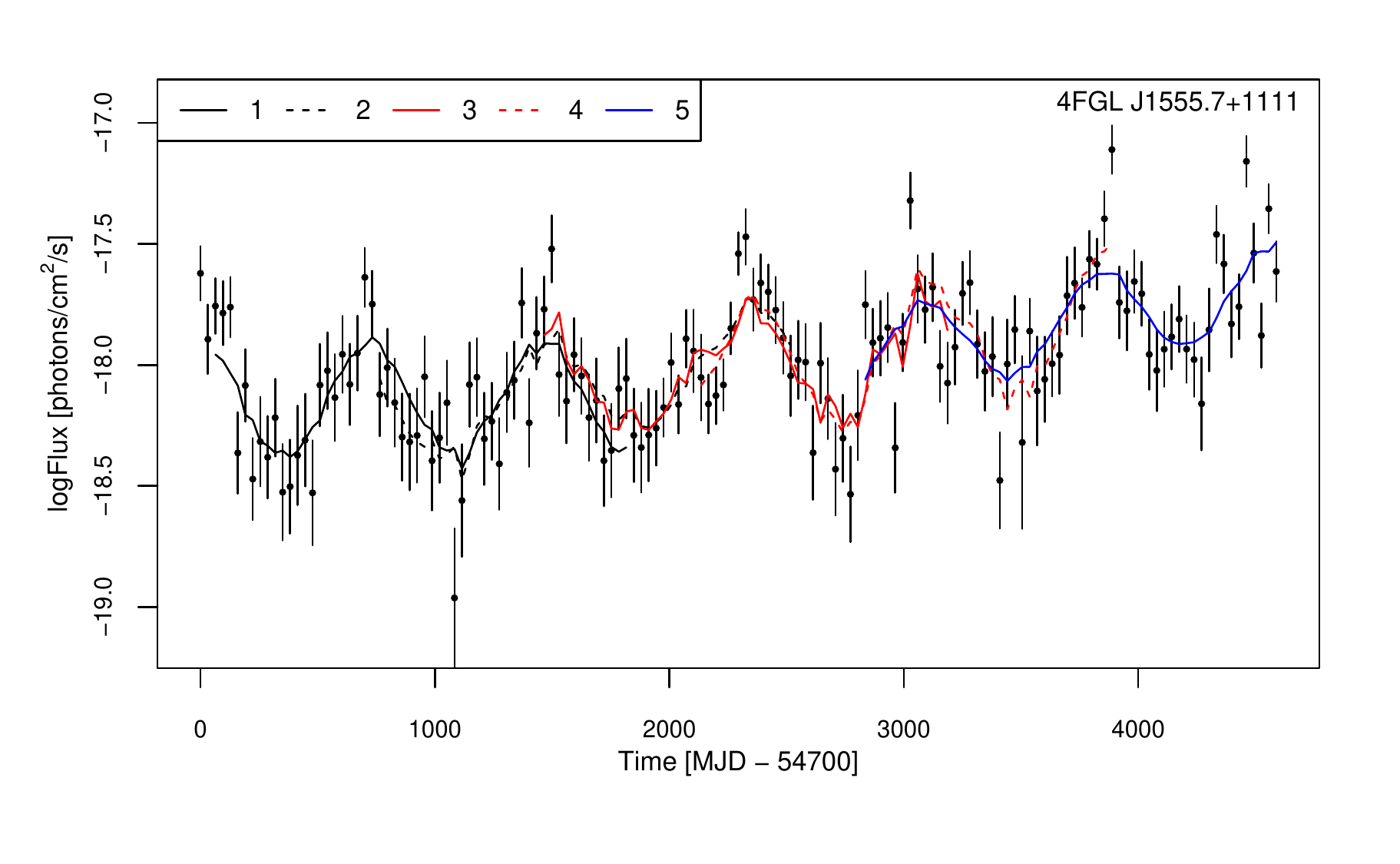}
\includegraphics[width = 0.45\textwidth]{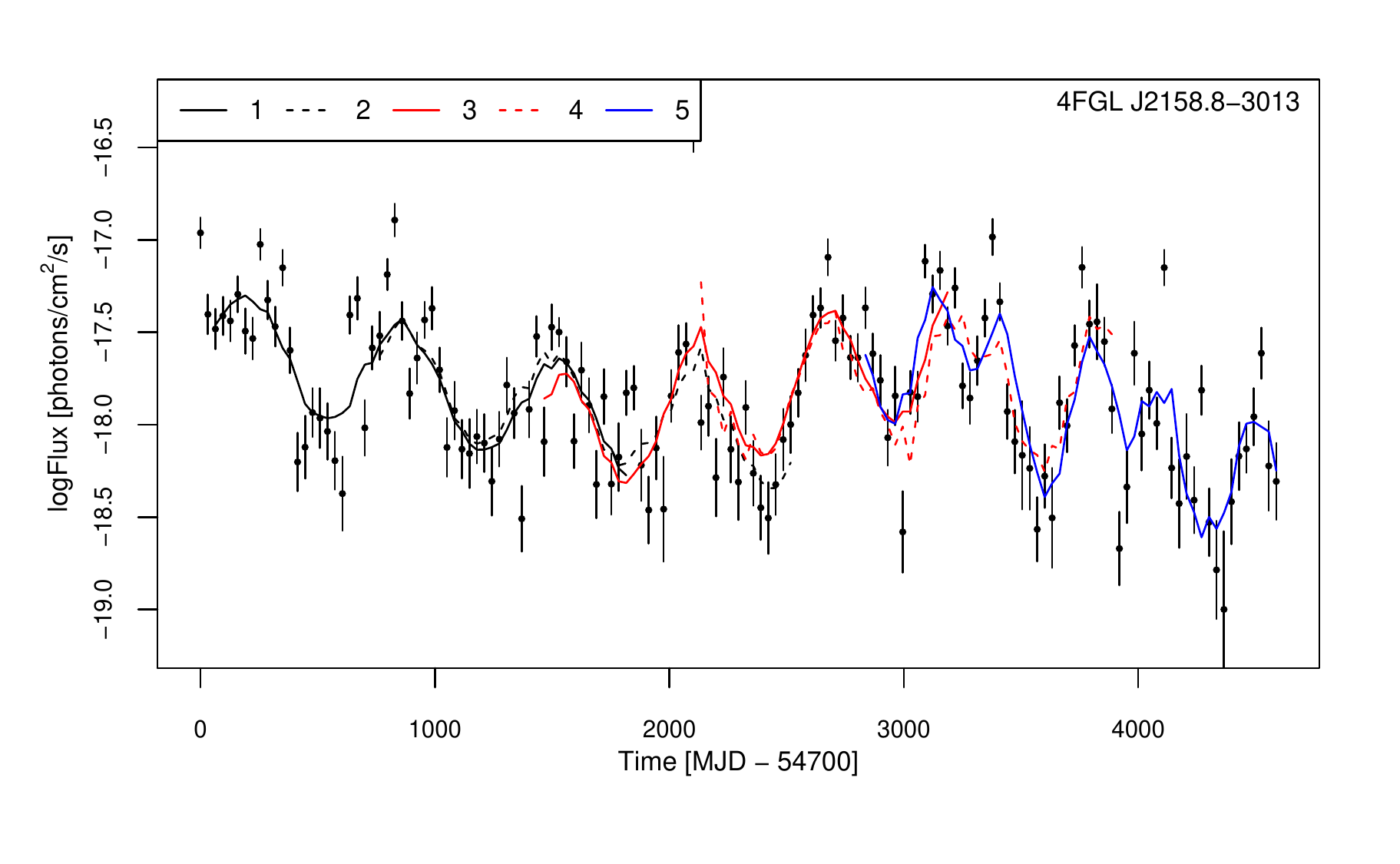}
\caption{MCMC fit for each time window. \label{fig:slice_fit}}
\end{figure}

For this analysis, several time windows are selected to search for periodicity change in the light-curves time span. 
Each time window has a width of 0.4 the full light-curve time span, this is, $0.4 \times 4620 \ \mathrm{days} \sim 1848 \ \mathrm{days} \ (\sim 5 \  \mathrm{ years})$. Thus, only the most significant periodic sources with periods smaller than $900 \ \text{days}$ are studied.
A total of 5 time windows are computed, each of them centred in 920, 1607, 2294, 2981, 3669 days from the start, respectively. 
The MCMC fits are applied at each time window for every source included. Then, an Agatha moving periodogram is performed as a cross-check as explained in section \ref{subsec:agatha}. The results are shown in Table \ref{tab:slices} and Figure \ref{fig:slice_p}. Figure \ref{fig:slice_fit}
 shows the light-curves analysed with the fitted MCMC components at each time window.
\begin{deluxetable*}{cccccccc|cc}
\tablenum{3}
\tablecaption{MCMC fit results and Agatha cross-check comparison of the AGN time windows.  For each source and window, the list indicates: the best model in terms of AIC; the AIC value; the period mean and standard deviation in days; the period 95\% Highest Density Intervals in days; the $\Delta$AIC between the periodic and the noise model; the p-value computed from $\Delta$AIC; the Agatha period mean and standard deviation: the Agatha lnBF. \label{tab:slices}}
\tablewidth{0pt}
\tablehead{
\colhead{4FGL Name} & \colhead{Window} & \colhead{Best Model} & \colhead{AIC} & \colhead{Period} & \colhead{Period HDI$_{95\%}$} & \colhead{$\Delta$AIC} & \colhead{p-value} & \colhead{Period\textsubscript{AGATHA}} & \colhead{lnBF}
}
\startdata
 & 1 & AR(1) + sin & 131.5 & 735 $\pm$ 32 & 671 - 798 & 14.32 & 1.5$\times 10^{-4}$ & 737 $\pm$ 74 & 4.7 \\
 & 2 & AR(1) + lin + sin & 125.26 & 838 $\pm$ 56 & 732 - 945 & 8.14 & 2.7$\times 10^{-3}$ & 851 $\pm$ 91 & 4.4 \\
J1555.7+1111 & 3 & AR(1) + lin + sin & 132.44 & 835 $\pm$ 74 & 701 - 984 & 5.39 & 9.8$\times 10^{-3}$ & 855 $\pm$ 118 & 3.8 \\
 & 4 & AR(1) + lin + sin & 128.86 & 758 $\pm$ 31 & 698 - 821 & 14.2 & 1.5$\times 10^{-4}$ & 772 $\pm$ 63 & 4.7 \\
 & 5 & AR(1) + lin + sin & 136.23 & 748 $\pm$ 61 & 651 - 858 & 9.96 & 1.2$\times 10^{-3}$ & 746 $\pm$ 98 & 1.5 \\ \cline{1-10}
 & 1 & AR(1) + lin + sin & 125.16 & 665 $\pm$ 34 & 596 - 732 & 12.78 & 3.0$\times 10^{-4}$ & 705 $\pm$ 88 & 3.3 \\ 
 & 2 & AR(1) + lin + sin & 142.25 & 617 $\pm$ 51 &  553 - 684 & 8.11 & 2.8$\times 10^{-3}$ & 623 $\pm$ 44 & 2\\
J2158.8-3013 & 3 & AR(1) + lin + sin & 134.96 & 564 $\pm$ 25 & 522 - 609 & 13.07 & 2.6$\times 10^{-4}$ & 575 $\pm$ 51 & 3.8\\
 & 4 & AR(1) + sin & 123.82 & 606 $\pm$ 57 & 535 - 669 & 7.51 & 3.7$\times 10^{-3}$ & 596 $\pm$ 43 & 3.9 \\
 & 5 & AR(1) + lin + harm & 130.46 & 678 $\pm$ 24 & 633 - 723 & 8.97 & 1.9$\times 10^{-3}$ & 688 $\pm$ 64 & -0.2 \\
\enddata
\end{deluxetable*}

For PG~1553+113, a periodic component is found in every time window with a p-value below $10^{-2}$. Windows 1, 4 and 5 present a period below the entire light-curve period of 774 days. For window 2 and 3, the period is $\sim 840$ days and with larger error bars. There is a total difference of 105 days between the minimum and the maximum period found. To quantify this dispersion, a constant fit is performed with a $\chi^2$ test, considering the standard deviations, that indicates the probability that the different time-window periods come from a constant distribution. The fit estimate of the constant value is $756 \pm 16 \ \mathrm{days}$, with a p-value 
equal to $0.64$. 
Agatha values agree with the MCMC results with a significance above $3.8$ for every time window except the last one with a value of $1.5$. 

For PKS~2155-304, the periodic components are also found with p-values below $10^{-2}$. Windows 1 and 2 present a period higher than the entire light-curve period of 613 days. Window 3 and 4 present a period below this value. For window 5, the best fitted model corresponds to an AR(1) stochastic component, a linear trend and a sinusoidal with a second harmonic component.
Again, the $\chi^2$ test is computed, resulting in a constant value of $618 \pm 25 \ \mathrm{days}$ with a p-value of $0.02$.

\begin{figure}[t!]
\centering
\includegraphics[width = 0.45\textwidth]{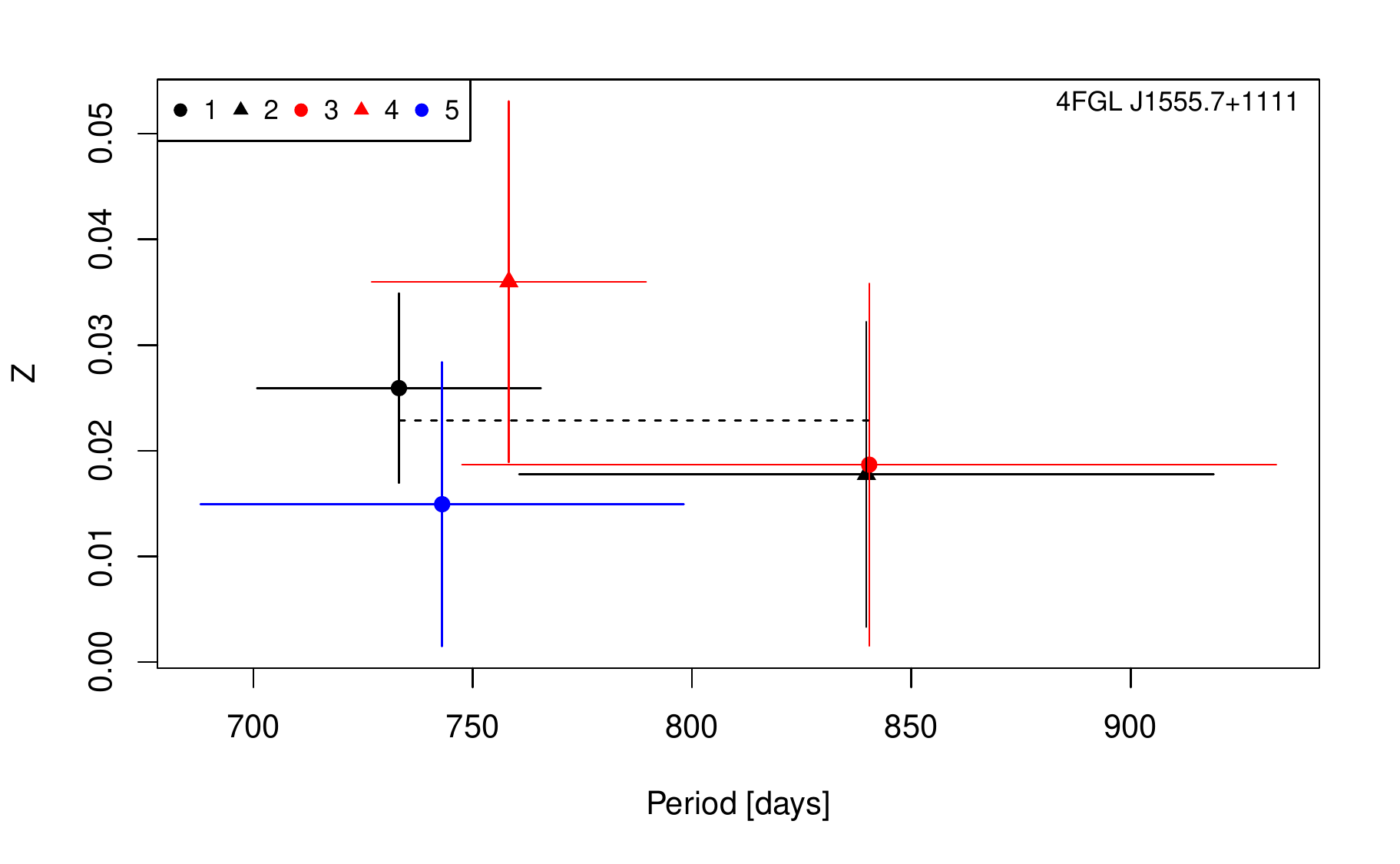}
\includegraphics[width = 0.45\textwidth]{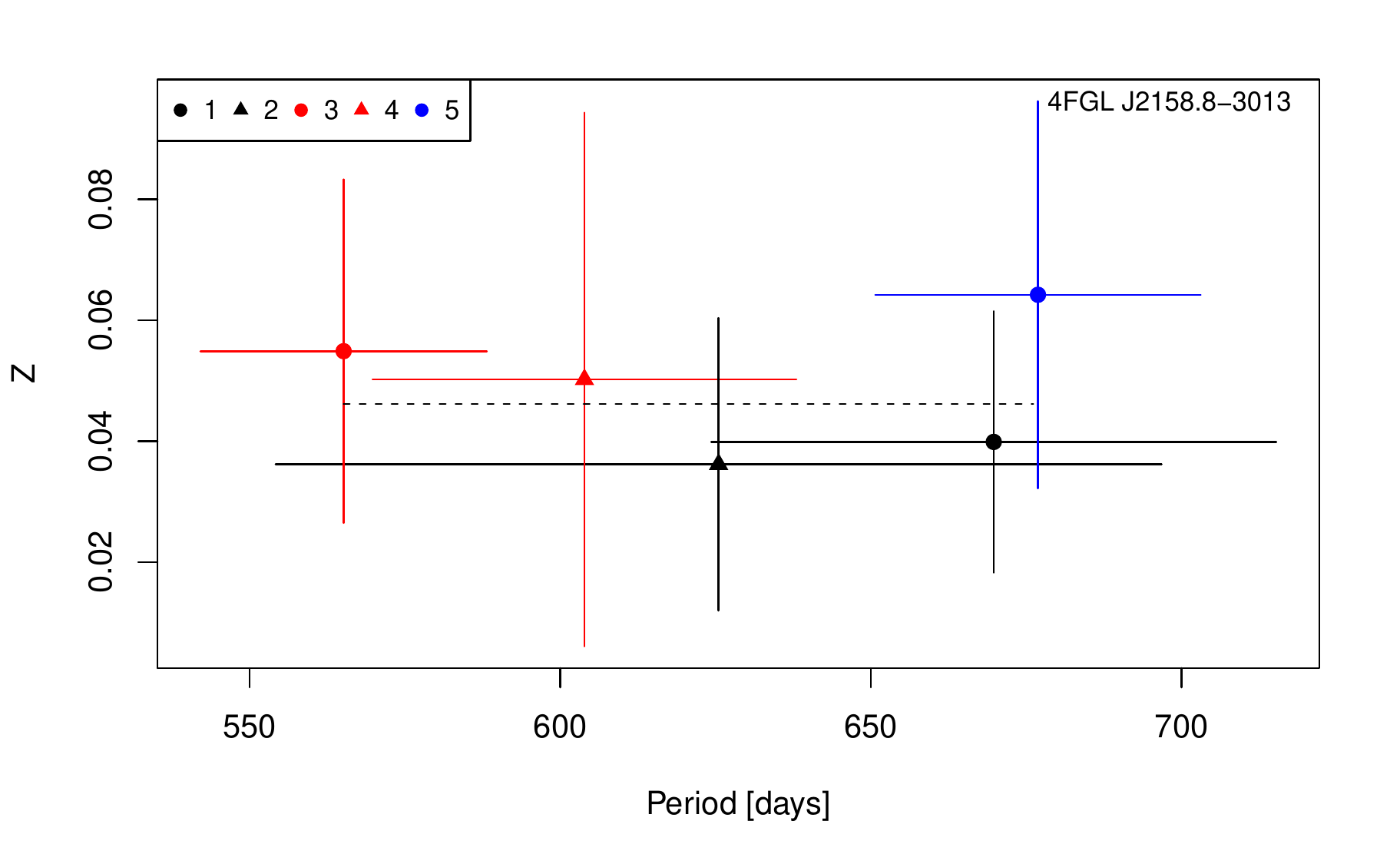}
\caption{MCMC physical amplitude Z mean and standard deviation for each time-window period. Horizontal dashed line indicates the result of the constant fit value.\label{fig:slice_A}}
\end{figure}

As explained in the Appendix \ref{app}, the transformed amplitude $Z'$ of the periodic modulations is obtained from the MCMC posterior values of parameters $A \ \mathrm{and} \ B$. Then, the physical amplitude $Z$ is derived from $Z'$ through $U$. Physical amplitudes as a function of the period are shown in figure \ref{fig:slice_A}. The $\chi^2$ test is computed to quantify the amplitude variation with period. This gives a p-value of $0.87$ and $0.95$ for PG~1553+113 and PKS~2155-304 respectively.

\section{Discussion and conclusions}
\label{sec:discussion}
Models explaining the periodicity of high-energy emission of quasars involve a variety of mechanism from jet precession (\cite{2013MNRAS.428..280C}) to periodic changes in the disk accretion flow \citep{2003MNRAS.344..468G} somehow transmitted to the jet.

An important class of models are geometric models
\citep{2004ApJ...615L...5R}. In geometric models, the periodicity in the observed emission is due to a change in the viewing angle of the jet components. 
The lighthouse model of \cite{1992A&A...255...59C} explains QPOs with periods of a few hundred days by the rotation of plasma bubbles around the central axis of the jet. In that model, the observed period increases with time and the amplitude of the emission is also time-dependent (see Fig. 2 and 3 of \cite{1992A&A...255...59C} ). 
Another possible mechanism  is based on the instability of the boundary (at transition radius $r_{tr}$) between an outer thin disk \citep{2003MNRAS.344..468G} or a torus \citep{2003MNRAS.341..832Z} and an inner radiatively inefficient flow (ADAF). In the model of \citet{2003MNRAS.344..468G}, the boundary is slowly moving outward and the period of oscillations increases with time. 

Finally, models with binary super-massive black holes (BSMBH) are natural candidates to explain an observed periodicity in the emission.
\citet{2017MNRAS.465..161S} have modelled the high-energy light-curve of PG~1553+113 with a geometrical model based on a BSMBH system in which one of the black hole has a precessing jet. In this model, the periodicity is due to the orbital motion of the BSMBH and not expected to change significantly with time while the flux amplitude is related to geometrical properties of the jet, which can change smoothly with time.
\citet{2018ApJ...854...11T} have interpreted the 2.2 year periodicity of the high energy light curve of PG~1553+113 with a model where blazar jets of a BSMBH are periodically perturbed by magneto-gravitational stresses \citep{2017ApJ...836..220C}. In this model, the smaller black hole stresses periodically the jet launched by the heavier one, triggering synchrotron emission and inverse Compton scattering in the GeV energy range. The observed emission could come either from a single jet or from the 2 jets of the BSMBH system. In the latter case, \citet{2018ApJ...854...11T} predict a stable period with smooth amplitude  changes from cycle to cycle, while in the former case the amplitude changes are erratic.

In this paper, a number of high-energy periodic source candidates from the Fermi-LAT 4FGL catalogue have been searched for periodicity by a novel method which separates clearly the stochastic and the periodic component in the light-curve fitting. Adding a periodic component to three sources, 4FGL~J1555.7+1111, 4FGL~J2158.8-3013 and 4FGL~J1903.2+5540 improves significantly the fit to the data compared to an ARMA (Auto-Regressive Moving Average) noise only model.  
A study of period and amplitude as a function of time was attempted by dividing the light-curve into different time windows for light-curves with periods less than 900 days. This excludes 4FGL~J1903.2+5540 from the time-window study. We now discuss the results of the time-window study for 4FGL~J1555.7+1111 and 4FGL~J2158.8-3013.

The measured period of the best candidate 4FGL~J1555.7+1111 does not significantly change with time. The amplitude of the periodic term depends only weakly on the period.
The almost constant amplitude and period are in agreement with a BSMBH model such as 
the 2-jet model of \citet{2018ApJ...854...11T}. 
An evolution of both the period and the amplitude are expected in models based on the lighthouse effect \citep{1992A&A...255...59C} such as the model of
\citet{2015ApJ...805...91M}. This model aims at explaining short term variability ($\le 1$ year), but could perhaps be extended to longer timescales \citep{2020A&A...634A.120A}.

In the case of 4FGL~J2158.8-3013, there is a marginally significant drift of the period with time. The amplitude of the oscillating component does not significantly change with time. The lack of correlation between amplitude and period disfavors again models based on the lighthouse effect. An harmonics of the period is detected in one of the time windows. This harmonics is not easily explained by pure stochastic noise (for instance by linear CARMA models).
The harmonics could be however the signature of oscillations in the disk of 4FGL~J2158.8-3013.  
These oscillations would trigger quasi-periodic accretion flows with harmonic frequencies and, by coupling between the disk and the jet, quasi-periodic variations in the observed flux.
The period of oscillation $P_{\mathrm{true}}$ at the source is related to the observed period  $P_{\mathrm{obs}}$ and to the source redshift $z$ by
 $P_{\mathrm{true}} = \frac{P_{\mathrm{obs}}}{(1+z)}.$
 For 4FGL~J2158.8-3013, one has $P_{\mathrm{true}} = 1.5$ yr.
If $M_{bl}$ is the black hole mass, $r_g = \frac{GM_{bl}}{c^2}$ is the gravitational radius, 
the transition from a disk or a torus to an ADAF occurs at radius
\begin{equation}
r_{tr} =  Kr_g \left( (\frac{10^{8} M_{\odot}}{M_{bl}})(\frac{P_{\mathrm{true}}}{1\ \mathrm{yr}})
\right)^{2/3}
\end{equation}
with $K \simeq 524$ for the transition radius to a disk
\citep{2003MNRAS.344..468G}
and $K \simeq 2100$ for the transition to a torus \citep{2006ApJ...650..749L}.

The estimates of the 4FGL~J2158.8-3013 black hole mass have a very large spread \citep{2010A&A...520A..23R,2008arXiv0810.1055D,2007ApJ...664L..71A}. Taking $M_{bl} = 10^{8} M_{\odot}$ as a typical value, one finds $r_{tr} \simeq 690 r_g$ in the disk model and $r_{tr} \simeq 2800 r_g$ in the torus model. The value of $r_{tr}$ for the disk model is too large according to \cite{2020A&A...634A.120A}. The value of $r_{tr}$ for the torus model is similar to the value obtained by \cite{2006ApJ...650..749L} for BL Lac
AO~0235+164.

The discussion on 4FGL~J1555.7+1111 and 4FGL~J2158.8-3013 has shown the advantages of our time-domain approach: the possibility of separating the periodic signal from the stochastic noise, to account for harmonics and to study the evolution in time of periods and amplitudes. This paper dealt exclusively with regularly sampled light curves. In the next step, we will extend the study to the whole Fermi-LAT data-set, including light-curves with holes in the observations.

\appendix
\section{Derivation of the MCMC model for regularly spaced data \label{app}}
Subtracting $k$ times equation \ref{eq:4_1}, one gets 
\begin{eqnarray*}
z &= \phi(t_n) -  \sum_{j=1}^{k} \beta_j \phi(t_{n-j}) \\
&= (1 - \sum_{j=1}^{k} \beta_j+\sum_{j=1}^{k} \beta_j j \delta t)\bar{\phi} +Ct_n(1 - \sum_{j=1}^{k} \beta_j) +w(\delta t) + S_n\\ 
S_n &= \sum_{j} (A_j (\cos(\omega_j t_n)-\sum_{l=1}^{k} \beta_l \cos(\omega_j t_{n-l}))+ B_j (\sin(\omega_j t_n)-\sum_{l=1}^{k} \beta_l \sin(\omega_j t_{n-l})) ) 
\end{eqnarray*}
with $\delta t = t_1-t_0.$

$S_n$ can be simplified by defining
\begin{equation}
    U_j\exp{i\psi_j} = 1 - \sum_{l=1}^{k} \beta_l \exp{(-i l \omega_j \delta t)}
\end{equation}
with $U_j$ real.

Then $$ S_n =\sum_{j} (A_j U_j \cos(\omega_j t_n+\psi_j)+ B_j U_j \sin(\omega_j t_n+ \psi_j)) $$
Finally, z can be written as
\begin{equation}
    z = \bar{\phi'}+\sum_{j} ({A'}_j \cos(\omega_j t_n) + {B'}_j \sin(\omega_j t_n)) + C't_n + w(\delta t) 
\end{equation}
with 
\begin{eqnarray}
\bar{\phi'} &= (1 - \sum_{j=1}^{k} \beta_j +\sum_{j=1}^{k} \beta_j j \delta t )\bar{\phi}\\
C'& = C(1 - \sum_{j=1}^{k} \beta_j) \\
A'_j&= U_j(A_j \cos(\psi_j) + B_j \sin(\psi_j)) \\
B'_j& =U_j(-A_j \sin(\psi_j) + B_j \cos(\psi_j)) 
\end{eqnarray}

The time-averaged square amplitude of each oscillating term is 
\begin{eqnarray}
{Z'_j}^2 &= 1/2( {A'_j}^2+{B'_j}^2) \\
& = 1/2 (U_j)^2 ({A_j}^2+{B_j}^2) = (U_j)^2 {Z_j}^2
\end{eqnarray}    

The transformed amplitude $Z'$ picks up an additional  period dependence compared to the physical amplitude $Z.$ Specializing to the case of an AR(1) model with a single period $T=\frac{2\pi}{\omega}$, the ratio of the transformed amplitude to the physical amplitude is
\begin{equation}
\frac{Z'}{Z} = U = \sqrt{(1+\beta^2-2\beta \cos{(\frac{2\pi \delta t}{T})})}.
\end{equation}

If $\beta >0 $ as in the Ornstein-Uhlenbeck model (equation \ref{eq:o-u-i}), $\frac{Z'}{Z}$ is a decreasing function of $T$ in the limit of small sampling times 
$(\delta t/T \ll 1).$

\section{Systematics of the MCMC search \label{syst}}
\subsection{Period prior dependence \label{ap:prior}}

\begin{figure}[t!]
\centering
\includegraphics[width = 0.33\textwidth]{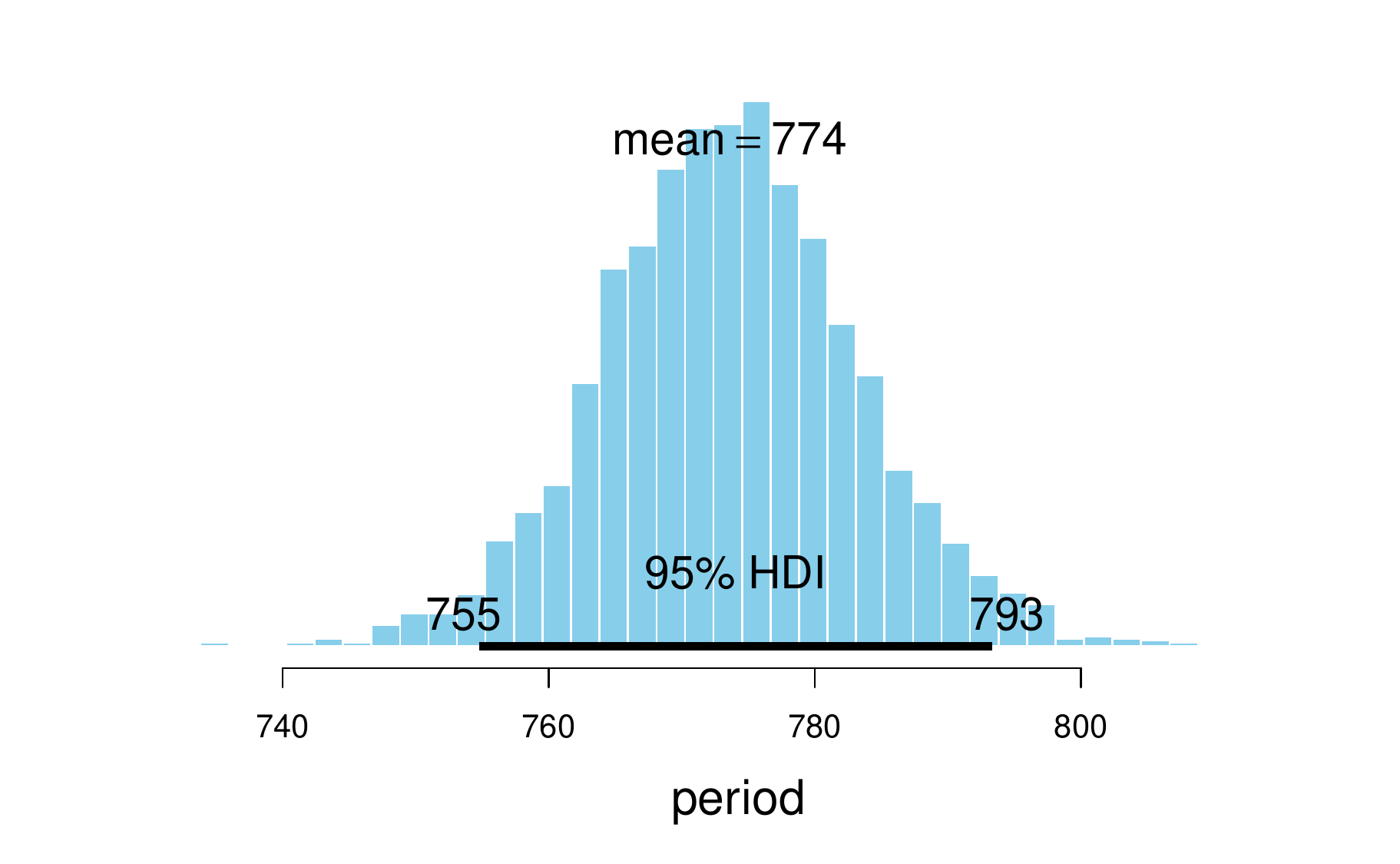}
\includegraphics[width = 0.33\textwidth]{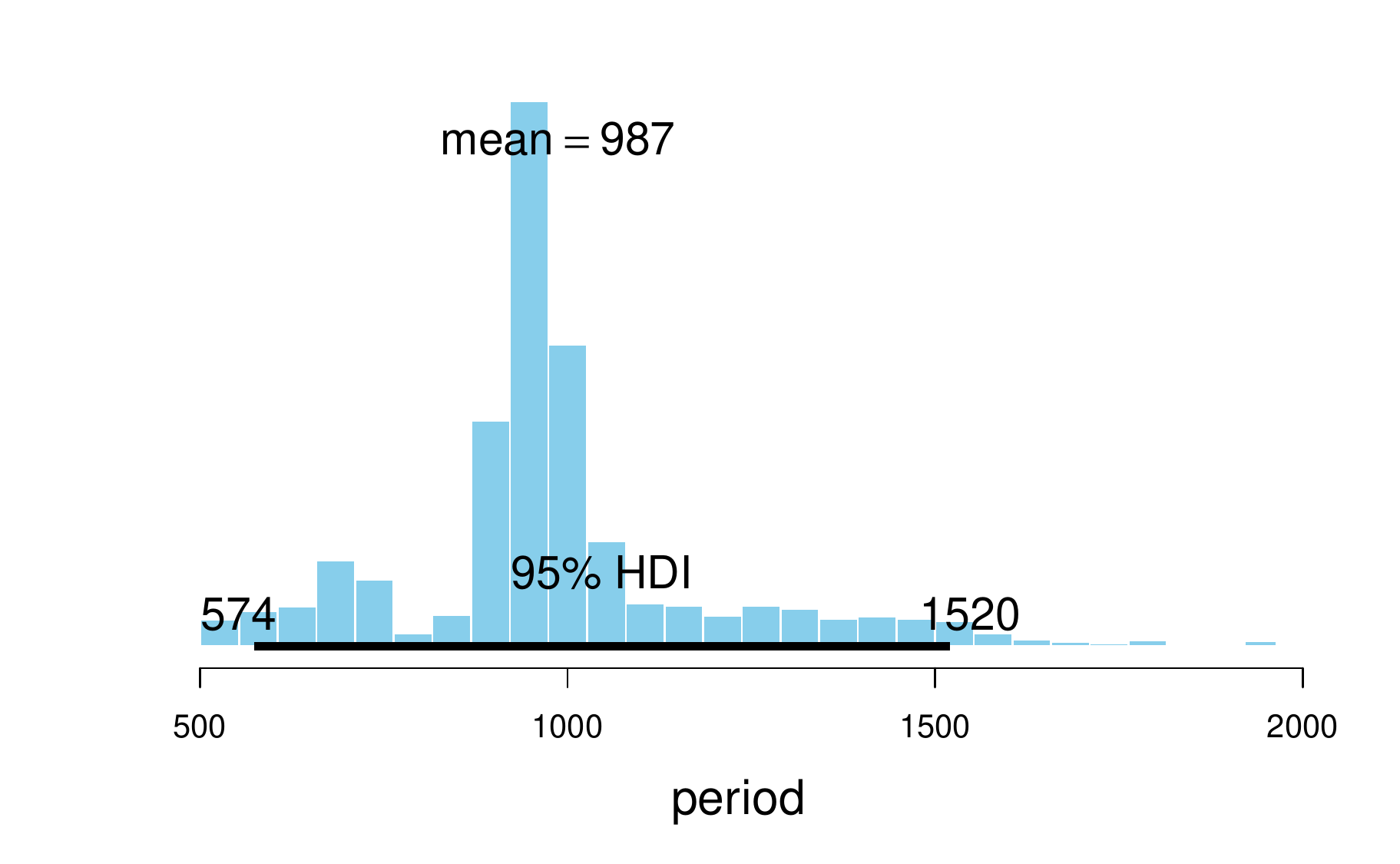}
\caption{ Examples of posterior distributions for the period parameter. [\textit{Left panel}] 4FGL J1555.7+1111  [\textit{Right panel}] 4FGL~J0721.9+7120. \label{fig:post}}
\end{figure}

On the high significance sources, changes in the prior distribution have no remarkable influence on the MCMC sampling and vague priors are suitable for the MCMC fit. The posterior distributions are approximately symmetric Gaussians from where the parameters are obtained as the mean value with standard deviation, as shown for example in the left panel of Figure \ref{fig:post} for PG~1553+113.

Right panel of Figure \ref{fig:post} shows an example of a poorer MCMC sampling where the chains are not converging properly and the posterior spreads wider to lower and higher values around the peak. 
Even so, the peak of the distribution is close to its mean value. 
This is the case for low significant sources marked with ** in Table \ref{tab:mcmc} where a change in the prior distribution does not change the output significantly.

\begin{figure}[t!]
\centering
\includegraphics[width = 0.33\textwidth]{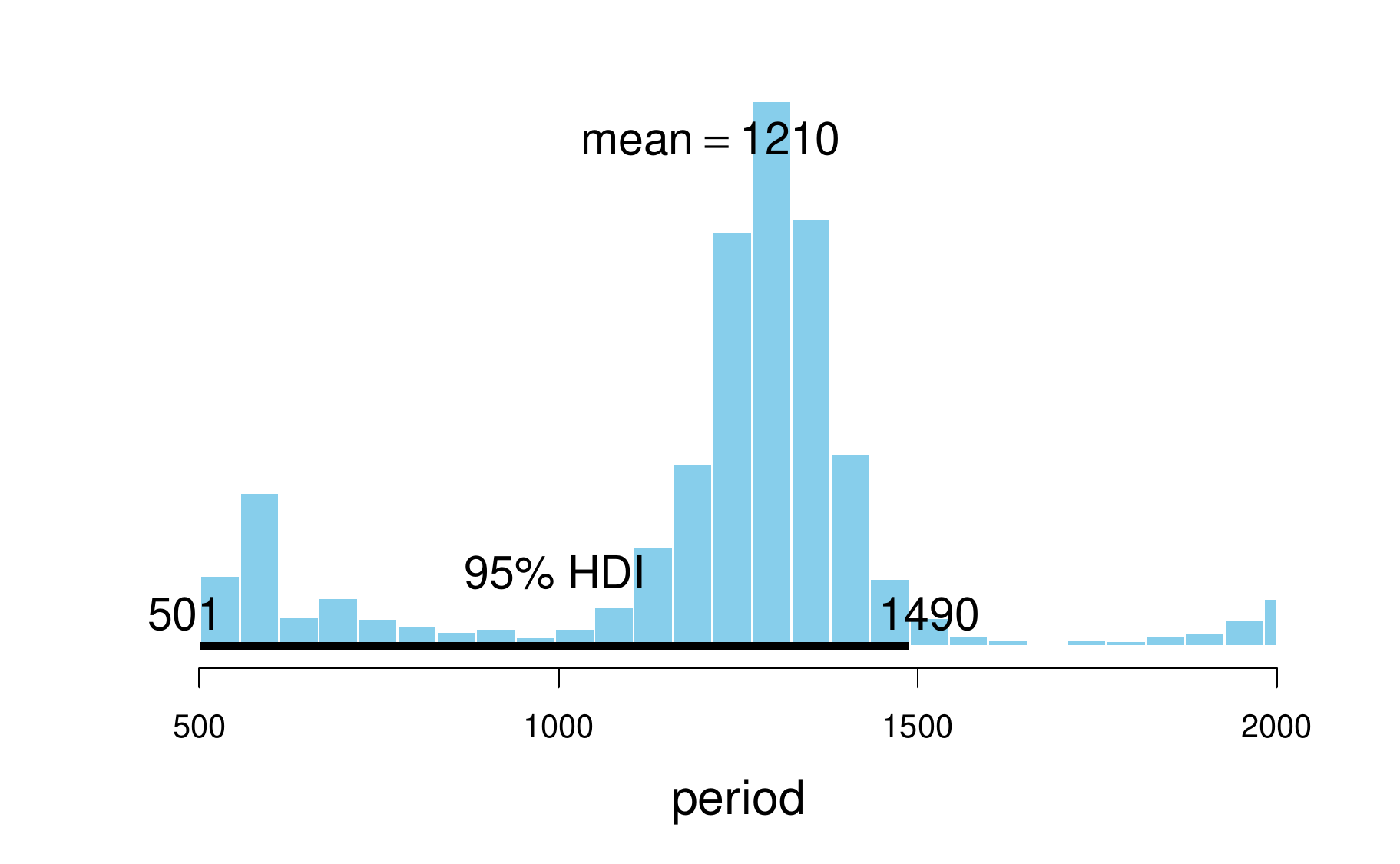}
\includegraphics[width = 0.33\textwidth]{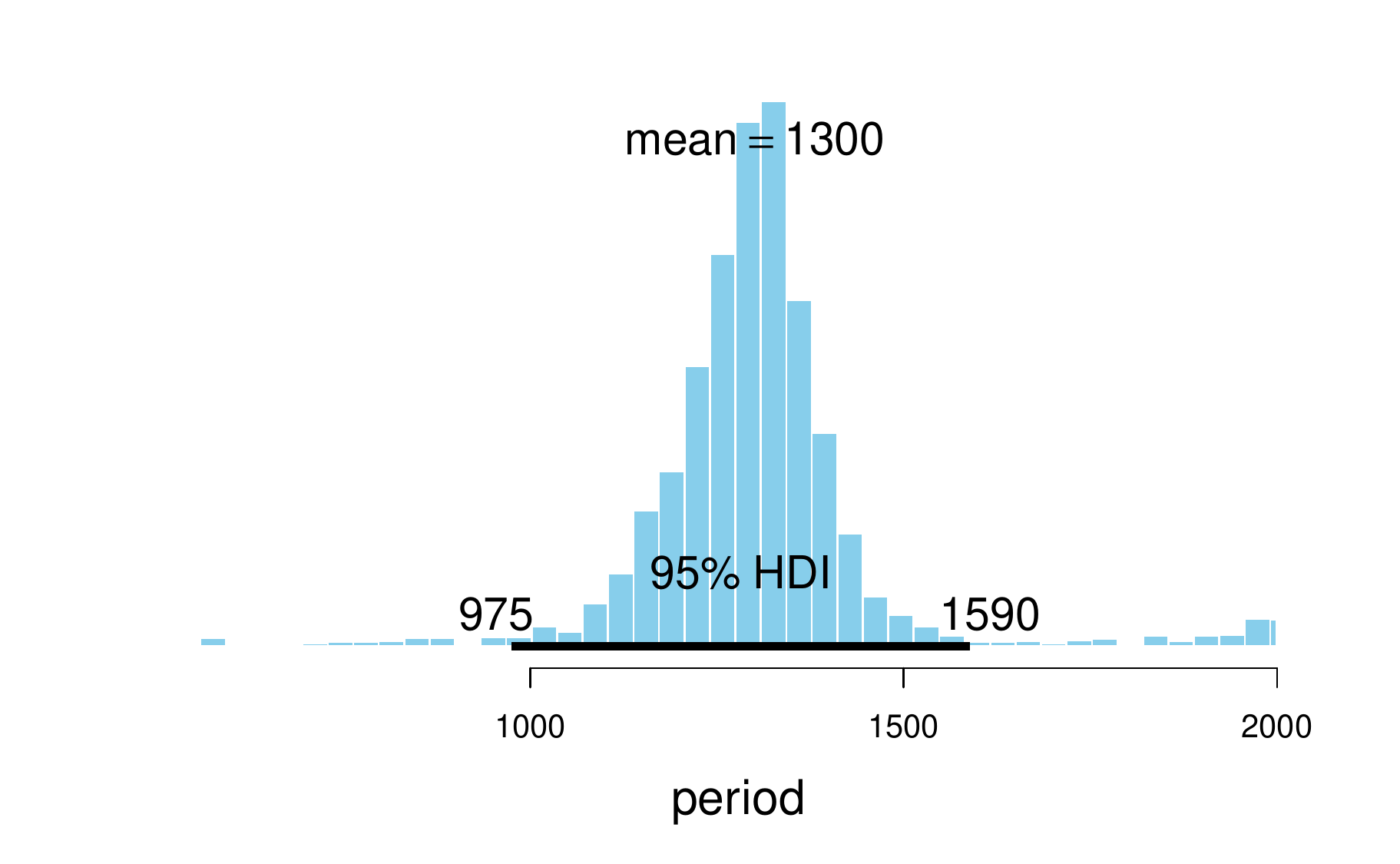}
\caption{ Examples of 4FGL J0457.0-2324 posterior distributions for the period parameter using: [\textit{Left panel}] General prior [\textit{Right panel}] Specific prior. \label{fig:prior}}
\end{figure}

For low significant sources marked with *, different priors might result in better posterior results.  As can be seen in the example on Figure \ref{fig:prior}, the use of a general prior leads to a result with bad chain convergence. Thus, the HDI and standard deviation are larger and the mean of the distribution is not close to the peak value. 
Centering the prior distribution in the highest value and reducing its standard deviation results in a posterior distribution closer to a symmetric Gaussian.
This shows a strong dependence between the posterior and the choice of the prior distribution. Thus, the acceptance of these results is lower and can be correlated with the inferior significance. 

\subsection{Correlations between parameters  \label{ap:post}}

\begin{figure}[t!]
\centering
\includegraphics[width = 0.9\textwidth]{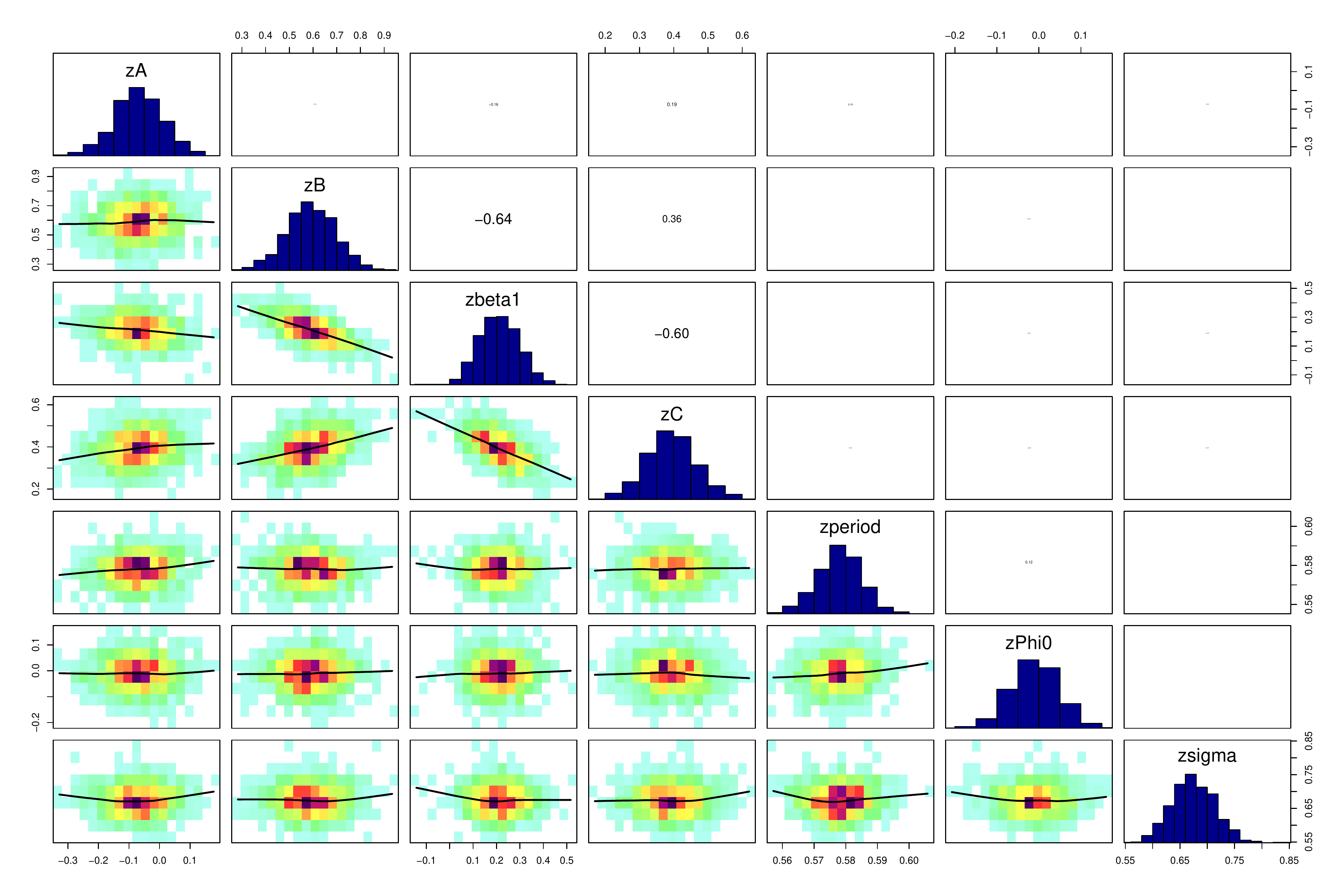}
\caption{ Example of a corner plot for 4FGL~J1555.7+1111. The diagonal plots show the posterior distributions for every standardize parameter. The 2D maps below the diagonal show the density of the posterior distribution with one parameter in each axis. The values above the diagonal show the Pearson correlation coefficients between parameters.  \label{fig:corr}}
\end{figure}

For every source analysed, the correlation between MCMC parameters is not as important as to affect the efficiency of the sampling chains. There are some correlations between periodic parameters (period, A and B) and between AR parameters ($\beta_1$, $\beta_2$) with the others in the model. All correlation values are below $\sim 0.6$.
In the Figure \ref{fig:corr} the corner plot for PG~1553+113 is shown, useful to visualize the pairwise correlations between model parameters.

\subsection{Tests for normality}
\begin{figure}[t!]
\centering
\includegraphics[width = 0.3\textwidth]{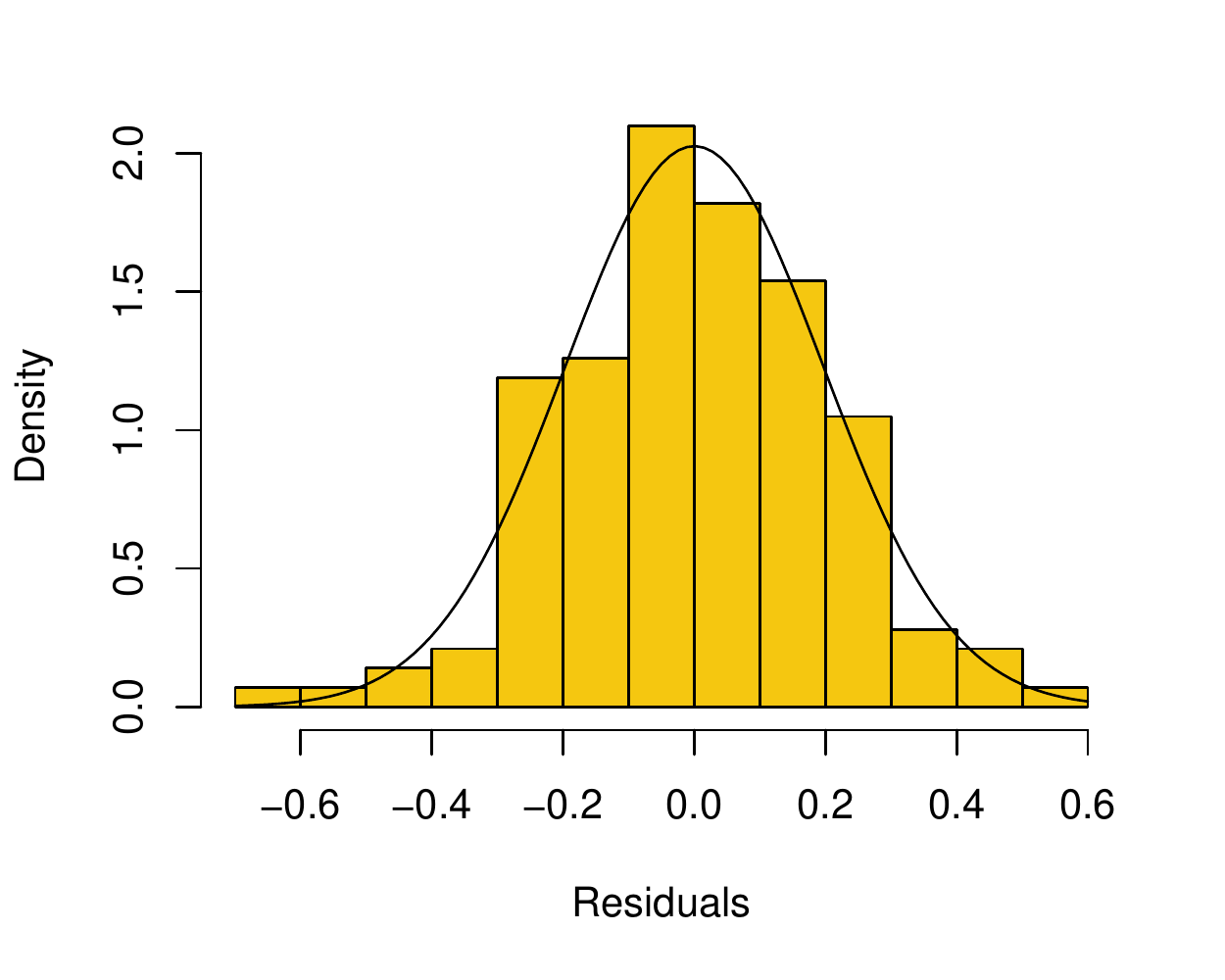}
\includegraphics[width = 0.3\textwidth]{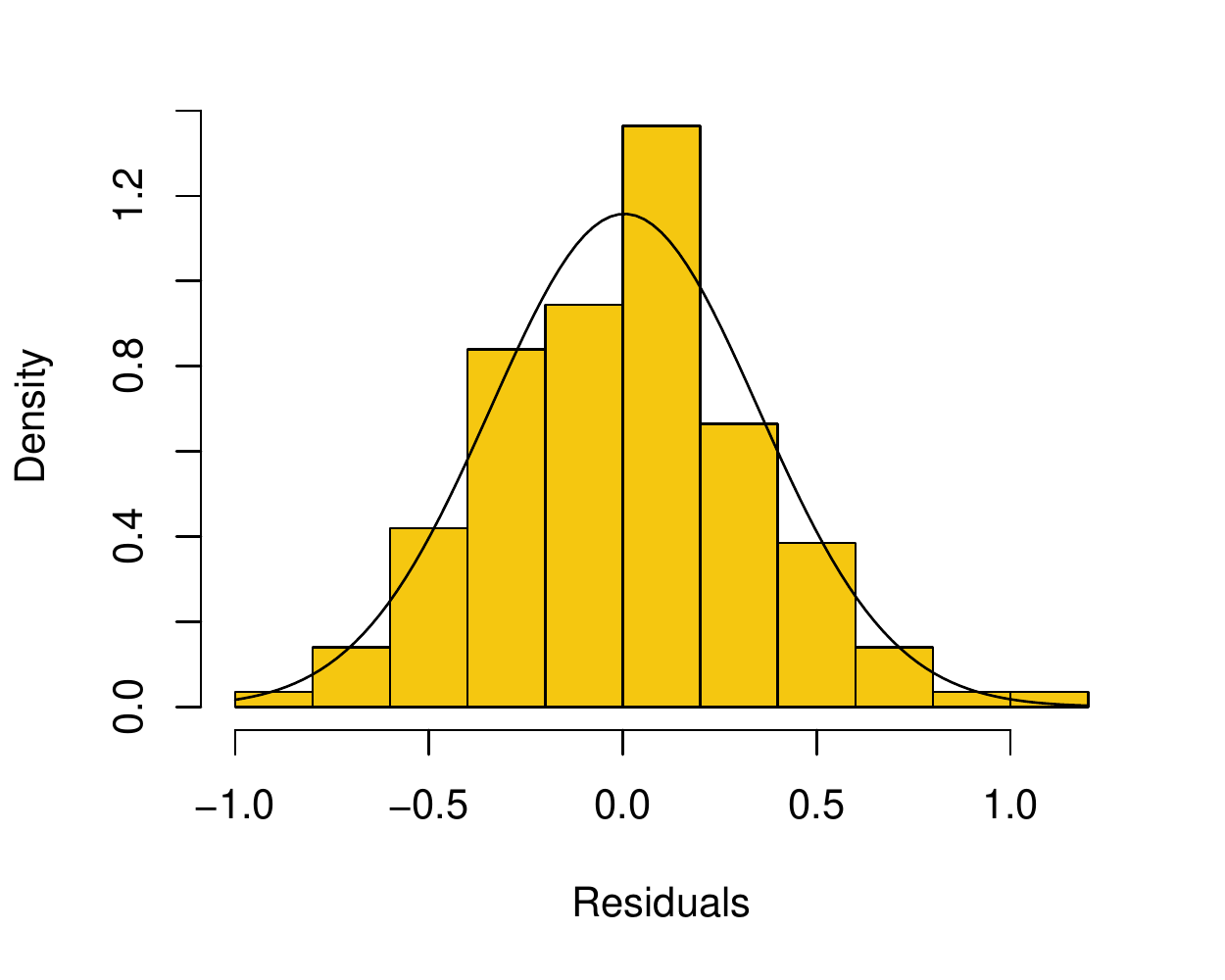}
\includegraphics[width = 0.3\textwidth]{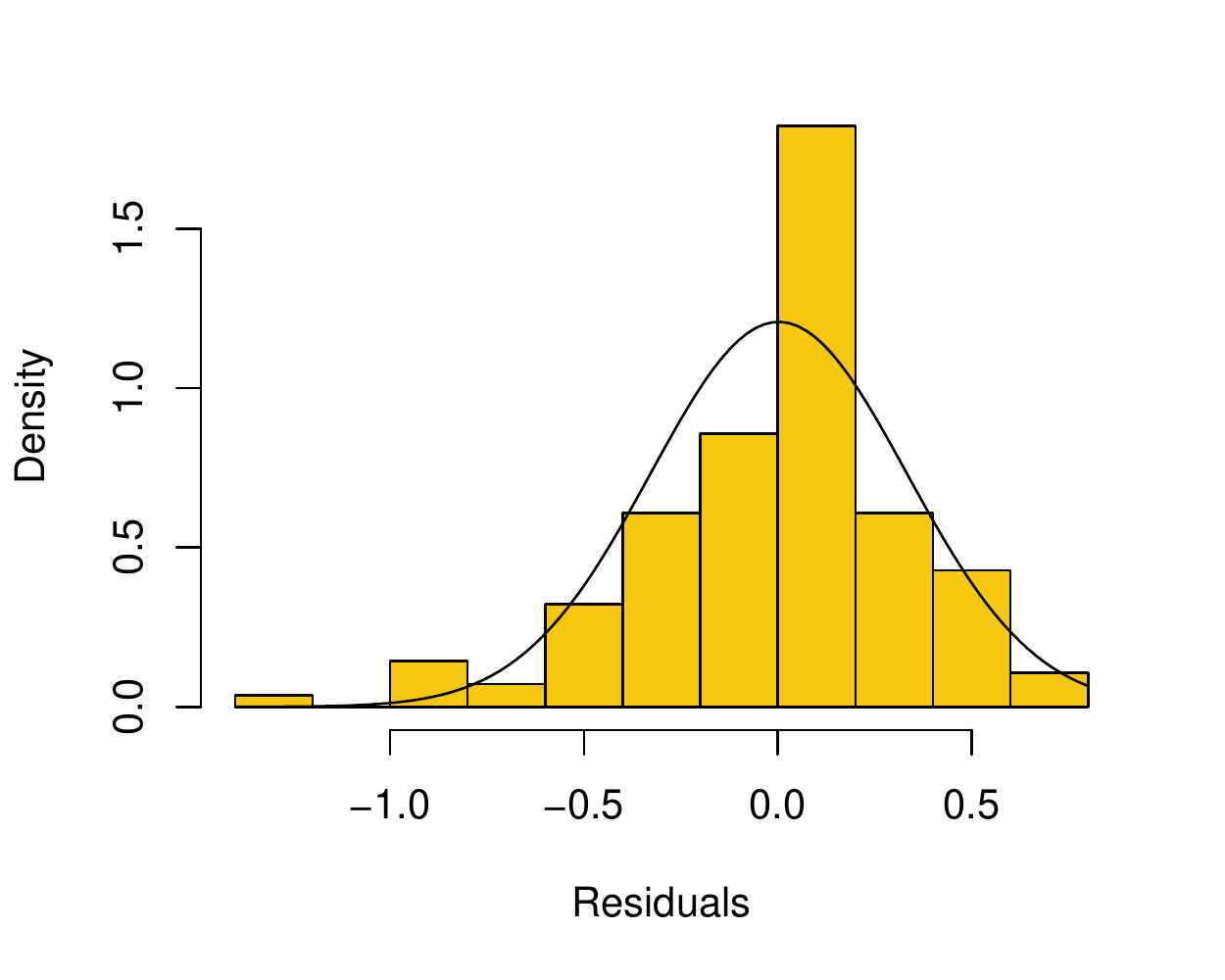}
\caption{ Residuals histogram for the three sources with most significant periodic signals. Black line shows a Gaussian distribution fit with $N_\text{mean}$ and $N_{\sigma}$ parameters given in Table \ref{tab:ad}. [\textit{Left panel}] 4FGL~J1555.7+1111 [\textit{Center panel}] 4FGL~2158.8-3013 [\textit{Right panel}] 4FGL~J1903.2+5540. \label{fig:ad}}
\end{figure}

As suggested by \citet{2018FrP.....6...80F} the results of the light curve fits have been tested for normality by performing an Anderson-Darling test on the residuals. Also, a Gaussian distribution $N(\text{mean}, \sigma)$ is fitted. The results are shown in Table \ref{tab:ad}. An example is shown in Figure \ref{fig:ad}.

The p-value form the Anderson-Darling test rejects the hypothesis of normality if its value is lower or equal to 0.05. For all sources but 4FGL 1903.2+5540 the p-value is above 0.05. As can be seen in Figure \ref{fig:ad} for 4FGL 1903.2+5540, some outliers are found in the left side of the residuals distribution and the test for normality fails.

\begin{figure}[t!]
\centering
\includegraphics[width = 0.32\textwidth]{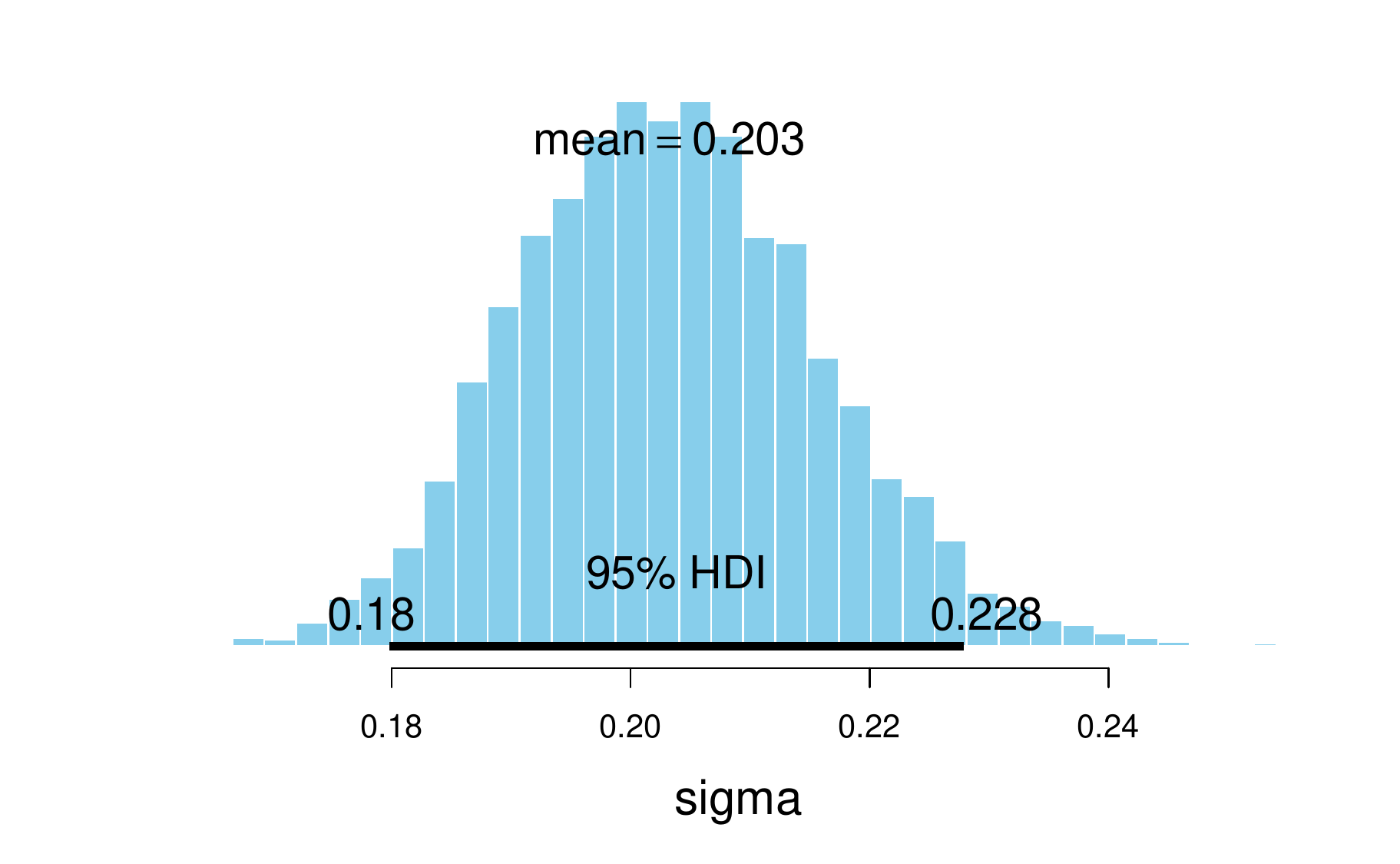}
\includegraphics[width = 0.32\textwidth]{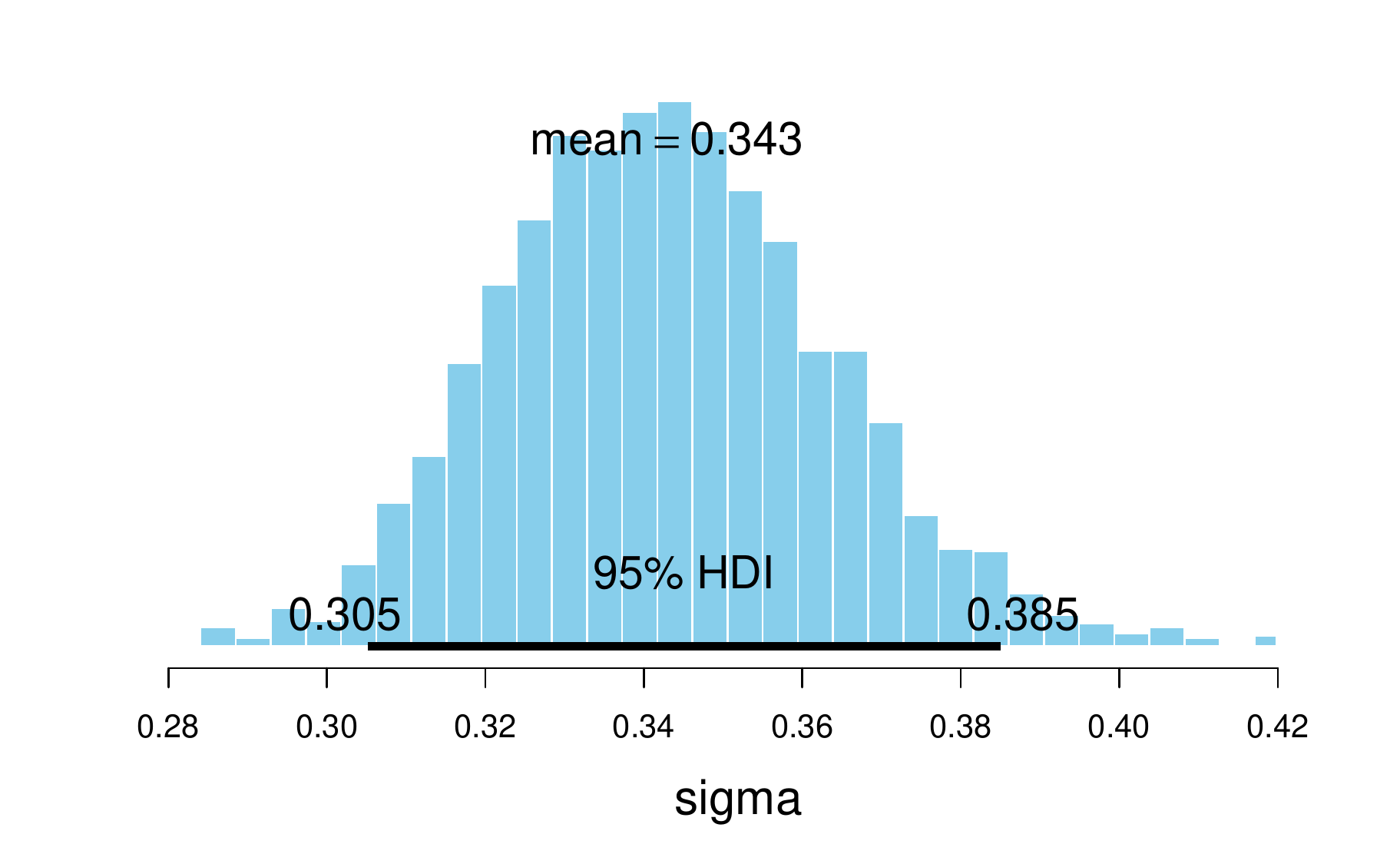}
\includegraphics[width = 0.32\textwidth]{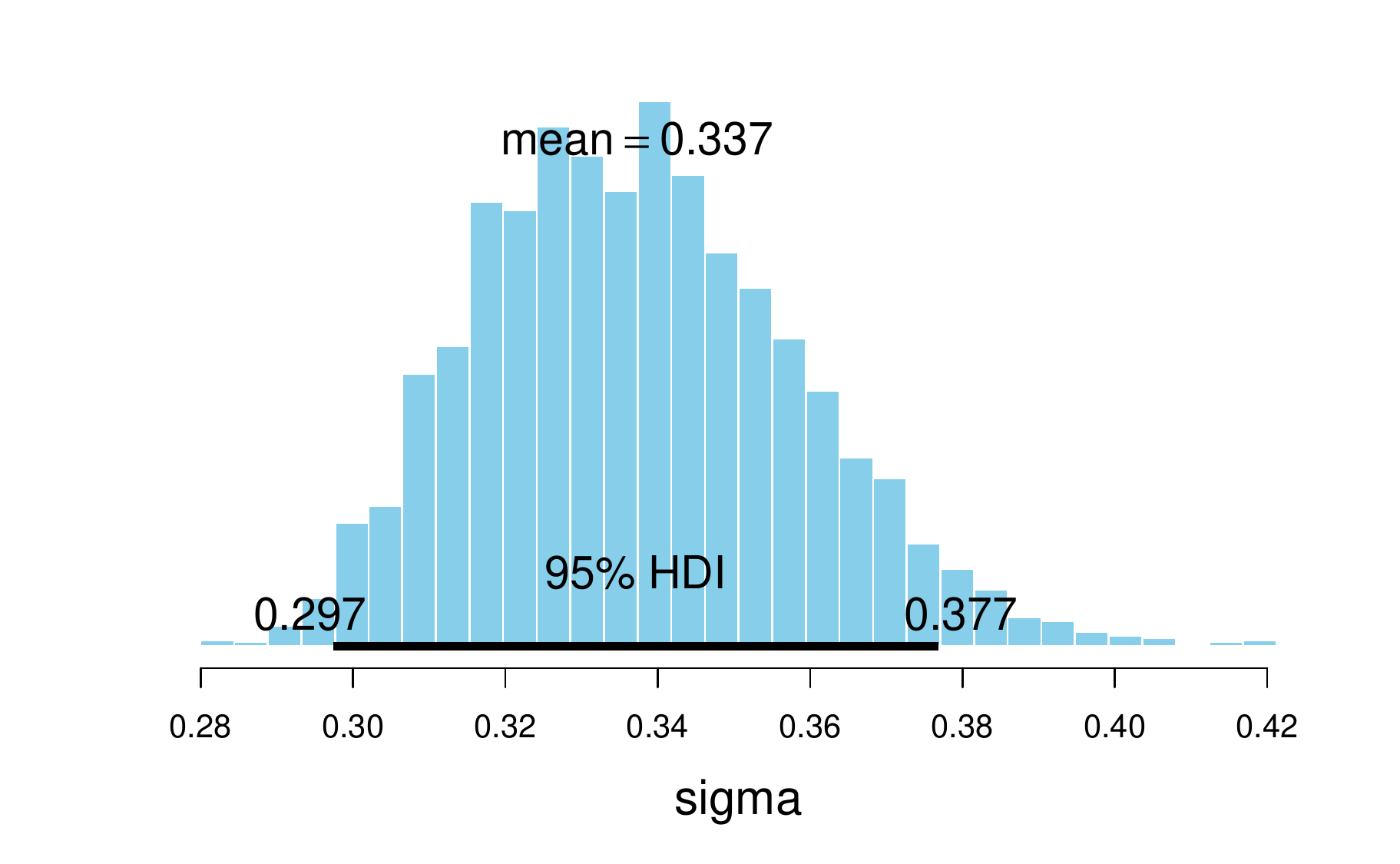}
\caption{ Examples of posterior distributions for the sigma parameter for the three sources with most significant periodic signals. [\textit{Left panel}] 4FGL~J1555.7+1111 [\textit{Center panel}] 4FGL~2158.8-3013 [\textit{Right panel}] 4FGL~J1903.2+5540. \label{fig:sigma}}
\end{figure}

From the Gaussian distribution fit some conclusions can be drawn. As expected from a proper data fit, the mean value of the residuals is centered close to 0. Furthermore, the standard deviation values $N_\sigma$ are equal to those obtained from the MCMC fit White Noise term $\sigma_\text{MCMC}$ and all posterior distributions are well sampled as approximately symmetric Gaussians. An example is shown in Figure~\ref{fig:sigma}.

\begin{deluxetable*}{ccccc}
\tablenum{4}
\tablecaption{ Test for normality of the AGN Fermi-LAT sample. For each source, the list indicates: the p-value of the Anderson-Darling test; Gaussian distribution fit mean $N_\text{mean}$ and standard deviation $N_{\sigma}$; the $\sigma$ of White Noise term in the MCMC fit. \label{tab:ad}}
\tablewidth{0pt}
\tablehead{
\colhead{4FGL Name} & \colhead{$p_{value AD}$} & \colhead{$N_{mean}$} & \colhead{$N_{\sigma}$} & \colhead{$\sigma_{\text{MCMC}}$}
}
\startdata
J1555.7+1111 & 0.89 & -1$\times 10^{-4}$ & 0.2 & 0.2  \\
J2158.8-3013 & 0.36 & 9$\times 10^{-4}$ & 0.33 & 0.34  \\
J1903.2+5540 & 9$\times 10^{-4}$ & -1.8$\times 10^{-3}$ & 0.33 & 0.34  \\
J0303.4-2407 & 0.38 & 7$\times 10^{-3}$ & 0.45 & 0.47  \\
J0521.7+2112 & 0.66 & 9.7$\times 10^{-3}$ & 0.41 & 0.43 \\
J1248.3+5820 & 0.43 & 5$\times 10^{-4}$ & 0.3 & 0.31 \\
J0211.2+1051 & 0.21 & 7.3$\times 10^{-3}$ &  0.46 & 0.46 \\
J0449.4-4350 & 0.29 &  1.4$\times 10^{-3}$ & 0.33 & 0.34 \\
J2202.7+4216 & 0.29 & -2.6$\times 10^{-3}$ & 0.53 & 0.55 \\
J0818.2+4222 & 0.324 & 8.7$\times 10^{-2}$ & 0.38 & 0.38 \\
J0721.9+7120 & 0.4 & 7$\times 10^{-4}$ &  0.52 & 0.53 \\
J0457.0-2324 & 0.3 & 1$\times 10^{-2}$ & 0.55 & 0.56 \\
J0428.6-3756 & 0.06 & 4$\times 10^{-3}$ & 0.51 & 0.52 \\
J0210.7-5101 & 0.86 & 7.8$\times 10^{-2}$ & 0.59 & 0.61
\enddata
\end{deluxetable*}

\bibliography{fermi_period}{}
\bibliographystyle{aasjournal}



\end{document}